\documentclass[aps,prb,twocolumn,showpacs,10pt,floatfix,superscriptaddress,floatfix]{revtex4-1}
\usepackage{amsmath}
\usepackage[normalem]{ulem}
\usepackage{ dsfont }
\usepackage{bm}             
\usepackage{mathtools}
\usepackage{amssymb}

\newcommand{\vk}{\vec k}
\newcommand{\ZZ}{\mathbb{Z}}
\newcommand{\vGamma}{\vec \Gamma}
\newcommand{\ve}{\vec e}

\newcommand{\vn}{\vec n}

\renewcommand{\vr}{\vec r}

\begin{document}

\title{Higher-order topological band structures}
\author{Luka Trifunovic}
\affiliation{Department of Physics, University of Zurich, Winterthurerstrasse 190, 8057 Zurich, Switzerland}
\author{Piet W. Brouwer}
\affiliation{Dahlem Center for Complex Quantum Systems and Physics Department, Freie Universit\"at Berlin, Arnimallee 14, 14195 Berlin, Germany}
\date{\today}

\begin{abstract}
The interplay of topology and symmetry in a material's band structure may
result in various patterns of topological states of different
dimensionality on the boundary of a crystal. The protection of these
``higher-order'' boundary states comes from topology, with constraints imposed
by symmetry. We review the bulk-boundary correspondence of topological
crystalline band structures, which relates the topology of the bulk band
structure to the pattern of the boundary states. Furthermore, recent advances
in the K-theoretic classification of topological crystalline band structures
are discussed.
\end{abstract}

\maketitle   

\section{Introduction}
Traditionally, the understanding of a material's band structure requires
knowledge of symmetry representation theory. During the last decades
it became increasingly clear that not only symmetries, but also
concepts borrowed from topology, and often a combination of the two, are
required for a complete understanding of band 
structures \cite{bernevig2013,ando2015,hasan2010,qi2011}. In
particular, gapped band structures can be classified into different
topological classes (also referred to as ``topological phases''), 
where two band structures
are in the same class if they can be smoothly deformed into
each other by changing system parameters, without closing the excitation 
gap and without reducing the symmetry of the band structure at an 
intermediate stage. This topological classification of band structures
applies
equally well to superconductors if these have a gapped excitation spectrum in
their BCS mean-field description.

An important practical consequence of the existence of distinct topological classes of band structures is that a nontrivial topology of the bulk band
structure may imply the existence of anomalous boundary states. A boundary
state is called ``anomalous'' if it cannot exist without the presence of the
topological bulk. Anomalous boundary states are immune to local perturbations
and can be removed only by a perturbation that closes the excitation gap of the
bulk band structure or reduces its symmetry. This connection between a
nontrivial topology of the bulk band structure and the existence of anomalous
boundary states is referred to as {\em bulk-bounda\-ry correspondence}. For
topological phases that are subject to non-spatial symmetries only, such as
time-reversal symmetry or the particle-hole antisymmetry of the superconducting
Bogoliubov-de Gennes Hamiltonian, the bulk-boundary correspondence is complete:
Each topological class of a $d$-dimensional bulk band structure is uniquely
associated with an anomalous boundary state of dimension $d-1$ and vice versa
\cite{kitaev2009,schnyder2009}.

\begin{figure}[t]%
\centering
\includegraphics*[width=0.9\columnwidth]{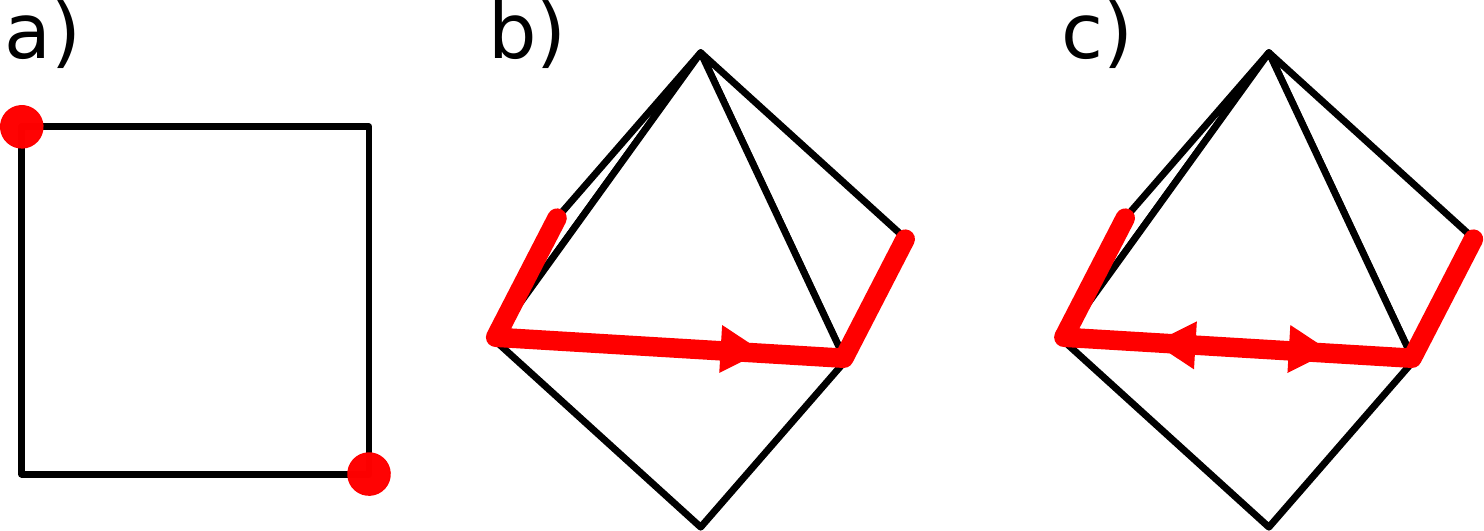}
\caption{Examples of higher-order boundary signatures: Majorana corner states of a two-dimensional topological crystalline superconductor (a), chiral hinge states of a three-dimensional second-order Chern insulator (b), or helical hinge states of a second-order topological insulator (c).}
\label{fig:0}
\end{figure}

The question of a bulk-boundary correspondence for topological phases that are
also subject to spatial symmetries, such as inversion or mirror symmetries, is
more subtle. In general, in this case the bulk-boundary correspondence is
incomplete and requires a degree of compatibility of the crystal termination
and the crystalline symmetries. Immediately after the discovery of topological
crystalline band structures
\cite{turner2012,fu2011}, it was understood that a conventional bulk-boundary
correspondence, in which the anomalous boundary states of a $d$-dimensional
crystal have dimension $d-1$, exists only for a boundary orientation that is
invariant under the action of the crystalline symmetry
group~\cite{ando2015,turner2012,fu2011,chiu2013,morimoto2013,shiozaki2014,trifunovic2017}.
This condition, however, can be met for a small number of symmetry groups only
--- mirror symmetry being an example ---, and even in those cases is restricted
to selected surface orientations. Recently, following pioneering work by
Schindler {\em et al.} \cite{schindler2018}, it was realized that topological
crystalline phases may also have anomalous boundary signatures of dimension
less than
$d-1$~\cite{schindler2018,benalcazar2017b,langbehn2017,song2017,fang2019,geier2018,khalaf2018,khalaf2018b,trifunovic2019,ahn2019,okuma2019,roberts2019,kooi2019,rasmussen2018},
provided the crystal termination as a whole respects the crystalline symmetry
group. Examples are anomalous states at hinges or corners of a
three-dimensional crystal or anomalous corner states of a two-dimensional
crystal, see Fig.~\ref{fig:0}. The condition that the crystal termination as a
whole respects the crystalline symmetry group is a much weaker condition on the
surface orientations than the condition that the orientation of individual
crystal faces is invariant under the crystalline symmetry. Moreover, it is a
condition that can be met for all crystalline symmetry groups. Topological
phases with this type of boundary signature are called {\em higher-order}
topological phases, where the order $n$ indicates the {\em codimension} of the
boundary states. In this terminology, topological phases that do not rely on
crystalline symmetries, which have boundary states of dimension $d-1$, are
called ``first-order''. 

Boundary states of codimension $n \ge 2$ may also occur as the anomalous
boundary states of nontrivial topological phase located on the crystal
boundary~\cite{langbehn2017,volovik2010,sitte2012,zhang2013}. Since such states
do not have their origin in the topology of the bulk band structure, they are
referred to as {\em extrinsic} \cite{geier2018}; anomalous boundary states that
are rooted in the topology of the bulk band structure are called {\em
intrinsic}. Although they are a property of the crystal termination, extrinsic
anomalous boundary states still have some degree of topological protection.
Specifically, extrinsic corner states cannot be removed by perturbations that
respect the crystalline symmetries and do not close the gaps along hinges or
surfaces of the crystal. Similarly, extrinsic hinge states cannot be removed by
symmetry-preserving perturbations that do not close surface gaps.

In this article we review the arguments that show how intrinsic higher-order
boundary states arise as a consequence of a topologically nontrivial band
structure. We discuss the simplified model systems that appeared in the
original publications and that have become paradigmatic examples of
higher-order topological band structures. We also discuss the
bulk-boundary correspondence for topological crystalline phases. Such
bulk-boundary correspondence was formulated by us for the case
of ``order-two'' crystalline symmetries that square to the identity, such as
mirror, inversion, or twofold rotation in 
Ref.~\onlinecite{trifunovic2019}. Here, these ideas
are extended to more general crystalline symmetry groups. Throughout the review
we restrict ourselves to ``strong'' topological phases, which are robust to a
breaking of the lattice translation symmetry (while preserving the crystalline
symmetries, of course).

We note that the relevance of topology to condensed matter systems is not only
limited to understanding band structures. Indeed, as the pioneering work of
Thouless, Kohmoto, Nightingale, and den Nijs shows \cite{Thouless1982}, 
the topological classes for quantized Hall systems can be defined even in 
the absence of a discrete translation symmetry. The results for strong
topological phases that we consider in this review, such as the bulk-boundary correspondence 
in presence of a crystalline symmetry, remain valid if the lattice translation
symmetry is broken in a manner that preserves the crystalline symmetries.
Some auxiliary results, such as the definition of topological invariants, 
rely on the existence of a band structure.

The remainder of this article is organized as follows: In Sec.~\ref{sec:1b} we
briefly review the ground rules for defining topological equivalence. In
Sec.~\ref{sec:2} we discuss three paradigmatic examples of topological band
structures in one, two, and three dimensions, their boundary signatures, and
how and under what conditions the existence of these boundary signatures is
rooted in the topology of the bulk band structure. Additionally, recent
experimental realizations of higher-order topology are reviewed. In
Sec.~\ref{sec:3} we formulate the formal bulk-boundary correspondence for
topological band structures with a crystalline symmetry. In Sec.~\ref{sec:5}
we discuss a specific example to make the rather general considerations of
Sec.~\ref{sec:3} more explicit. We conclude in Sec.~\ref{sec:6}.

\section{Topological equivalence} \label{sec:1b}

For a precise topological classification of band structures and the associated
boundary signatures, one has to define the ``rules'' of topological
equivalence. When taken literally, the definition of topological equivalence
stated in the first paragraph of the introduction implies that two band
structures with different numbers of occupied bands belong to different classes.
It is customary to relax this criterion and allow that the separate addition of
(topologically trivial) occupied or empty bands or sets of bands does not
change the topological class of the band structure. This topological
equivalence of band structures modulo the addition of trivial occupied or empty
bands is known as ``stable equivalence''. The complete bulk-boundary
correspondence for to\-pological band structures subject to non-spatial
symmetries was derived using the rules of stable equivalence
\cite{kitaev2009,schnyder2009}. 

More precisely, under the rules of stable equivalence one considers pairs ($H$,
$H$') of band structures with equal number of bands --- an approach known as
the {\em Grothendieck construction}. Two pairs $(H_1,H_1')$ and $(H_2,H_2')$
are considered topologically equivalent if $H_1 \oplus H_2'$ can be smoothly
deformed into $H_2 \oplus H_1'$, without closing the excitation gap and without
violating the symmetry constraints. With this definition, topological classes
acquire a group structure, the group operation being the direct sum of band
structures: Using the pair $(H,H')$ to represent its topological class, the
group operation is $(H_1,H_1') \oplus (H_2,H_2') = (H_1 \oplus H_2, H_1' \oplus
H_2')$. The inverse group operation is also defined: $(H_1,H_1') \ominus
(H_2,H_2') = (H_1 \oplus H_2',H_2 \oplus H_1')$. In this manner, the
topological phases are classified with an Abelian {\em classifying group} $K$.
The corresponding classification scheme is known as the 
``K-theory classification''. 

For completeness, we mention that the topological classification based on
stable equivalence is not the only type of classification used in the
literature. One example of a different classification is the ``non-stable''
classification of gapped band structures with a fixed number of bands. A well
studied classification of this kind concerns the Hopf
insulator~\cite{moore2008}. The Hopf insulator does not have a bulk-boundary
correspondence of the type mentioned above, although generalized bulk-boundary
correspondence was formulated recently~\cite{alexandradinata2019}. Like the
stable classification, the non-stable classification also has a group
structure, although the group operation can not be defined by superimposing two
band structures, since the number of bands is kept fixed. For band structures
defined on a $d$-dimensional sphere instead of the Brillouin zone, the group
structure is given by concatenation completely analogous to the group structure
of homotopy groups~\cite{hatcher2002}. Another alternative classification scheme
is one that allows for the addition of bands, but only if these are separated
by gaps from the existing bands. With such classification rules, the resulting
group structure of the classes can be non-Abelian~\cite{wu2019,bouhon2019}. The
``fragile'' topological classification scheme allows the addition of
non-occupied  bands or set of bands, but not for the addition of occupied bands
\cite{po2018}. As a result, the fragile topological classification results in a
monoid~\cite{song2019}, in which addition is defined as a direct sum, but
subtraction is not defined. A bulk-boundary correspondence has been formulated
for fragile topological phases~\cite{song2019b}, although its relevance to
condensed matter systems is unclear at the moment. Finally, one can define a
binary classification that only distinguishes between ``topologically trivial''
and ``topologically non-trivial'' \cite{bradlyn2017}, where a band structure is
called ``trivial'' if it is ``Wannierizable'', {\em i.e.}, it can be obtained
from localized states that respect the symmetry constraints. If only local
symmetries are imposed, this definition of topological
triviality~\cite{soluyanov2011,thonhauser2006,thouless1984} agrees with the
triviality of the stable equivalence classification. The two definitions differ
when non-local crystalline symmetries are imposed. In that case, Wannierizable
band structures are topologically connected to ``atomic-limit phases''. The
binary classification does not have a group structure, since superimposing two
``nontrivial'' band structures may or may not result in a ``trivial'' band
structure.

\section{Boundary signatures of topological band structures} \label{sec:2}
In this Section we first consider three examples of gapped band structures with
crystalline symmetries in dimensions $d=1$, $2$, and $3$. The examples are
meant to illustrate the role of the crystalline symmetries as well as the
non-spatial symmetries for the topological characterization of the band
structure and the protection of eventual boundary states. Subsequently, we
review recent experiments that realize higher-order topological band
strucutres. The last three subsections explain how and when the boundary
signatures derive from the nontrivial topology of the bulk band structure.

\subsection{One-dimensional model with inversion symmetry} \label{sec:3.1}
\begin{figure}[t]%
\centering
\includegraphics*[width=0.5\columnwidth]{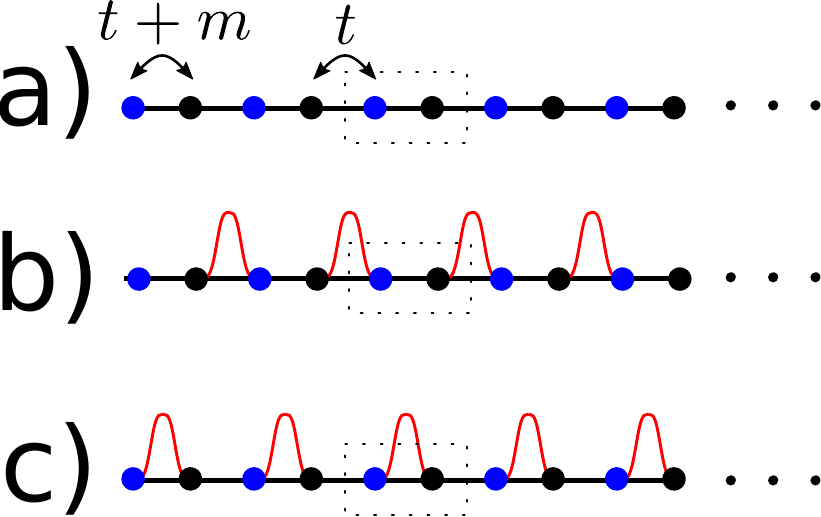}
\caption{Without particle-hole symmetry, the model (\ref{eq:Kitaev}) describes a one-dimensional lattice with a two-atom unit-cell and nearest-neighbor hopping amplitudes that alternate between $t$ and $t + m$ (a). In the ``topological phase'' $-2 t < m < 0$ Wannier functions are localized between unit-cells (b), whereas they are localized in the center of the unit-cells if $m < -2t$ or $m > 0$ (c). The two patterns can not be smoothly deformed into each other in the presence of inversion symmetry.}
\label{fig:1}
\end{figure}

As a first example, we consider the one-dimensional model known as the
Su-Schrieffer-Heeger model or the ``Kitaev chain'' in its non-superconducting
or superconducting realizations, respectively \cite{su1979,kitaev2001},
\begin{equation}
  H(k) = \Gamma_0 [m + t (1 - \cos k)] + \Gamma_1 t \sin k,
  \label{eq:Kitaev}
\end{equation}
with ``Dirac gamma matrices'' $\Gamma_0 = \sigma_1$ and $\Gamma_1 = \sigma_2$. The model is invariant under particle-hole conjugation ${\cal P}$,
\begin{equation}
  H(k) = -U_{\cal P} H(-k)^* U_{\cal P},
\end{equation}
with $U_{\cal P} = \sigma_3$. If ${\cal P}$ is not enforced,
Eq.~(\ref{eq:Kitaev}) represents a one-dimensional chain with a two-atom
unit-cell and nearest-neighbor hopping amplitudes that alternate between $t$
and $t + m$, see Fig.~\ref{fig:1}a. With particle-hole symmetry,
Eq.~(\ref{eq:Kitaev}) represents the Bogoliubov-de Gennes Hamiltonian of a
one-dimensional superconductor with a one-atom unit-cell, be it with a
non-standard form of the particle-hole conjugation operation. In addition, the
Hamiltonian (\ref{eq:Kitaev}) is also invariant under time-reversal ${\cal T}$,
the ``chiral antisymmetry'' $U_{\cal C}$, and inversion ${\cal I}$,
\begin{align}
  H(k)  =&\, U_{\cal T} H(-k)^* U_{\cal T} \nonumber \\
  =&\, -U_{\cal C} H(k) U_{\cal C} \nonumber \\
  =&\, U_{\cal I} H(-k) U_{\cal I},
\end{align}
with $U_{\cal T} = 1$, $U_{\cal C} = U_{\cal P} U_{\cal T} = \sigma_3$, and
$U_{\cal I} = \Gamma_0 = \sigma_1$. The inversion operation ${\cal I}$ commutes
with ${\cal T}$, but it anticommutes with ${\cal P}$, so that this model
represents an odd-parity superconductor if ${\cal P}$ is present. 

The model (\ref{eq:Kitaev}) describes a one-parameter family of Hamiltonians
$H(k)$, labeled by the parameter $m$. The spectral gap of the model
(\ref{eq:Kitaev}) closes at $m=0$. To see, whether this gap closing point
represents a topological phase transition, one looks for the existence of
``mass terms'', perturbations to $H(k)$ that anticommute with the matrices
$\Gamma_0$ and $\Gamma_1$. If such mass terms do not exist, a gap closing can
not be avoided when $m$ is tuned through the gapless point at $m=0$ and the
gapless point represents a topological phase transition. If it exists, the
phases below and above $m=0$ are topologically equivalent.

By inspection, one easily verifies that the model (\ref{eq:Kitaev}) allows a
single mass term, $M = \sigma_3$.  This mass term is, however,
incompatible with ${\cal P}$, ${\cal C}$, or ${\cal I}$, so that the gapless
point at $m=0$ represents a topological phase transitions if at least one of
these three symmetries is present. With ${\cal P}$ or ${\cal C}$ the model
(\ref{eq:Kitaev}) is in a first-order ({\em i.e.}, non-crystalline) topological
phase for $-2 t < m < 0$ with a (Majorana) zero mode at each end. Without
${\cal P}$ or ${\cal C}$, but with ${\cal I}$, the model no longer has
protected zero-energy end states, but it has an anomalous half-integer ``end
charge'' if $-2t < m < 0$. The existence of the end charge follows from a
calculation of the bulk polarization \cite{lau2016}. It can also be inferred
from the fact that the mid-gap end state in the presence of ${\cal
P}$ or ${\cal C}$ symmetrically
removes a half-integer charge from the valence and conduction
bands. If ${\cal P}$ or ${\cal C}$ is broken, the end states can be removed,
{\em e.g.}, by a local potential at each end, but the half-integer end charge
is immune to the addition of a local perturbation,
since the inversion symmetry ${\cal I}$ prevents charge flow through the insulating bulk. If ${\cal P}$, ${\cal C}$, and ${\cal I}$ are broken the existence of the mass term $M = \sigma_3$ implies that the model is topologically trivial.

Having no anomalous end states, the model (\ref{eq:Kitaev}) with broken ${\cal
P}$ or ${\cal C}$ symmetry is an example of a ``Wannierizable'' or ``atomic
limit'' band structure, a band structure for which there exists a basis of
localized ``Wannier functions''. That such atomic-limit band structures can
nevertheless be topologically distinct can be seen by inspecting the model
(\ref{eq:Kitaev}) in the limits $t/m=-1$ and $t/m = 0$ in which the system
dimerizes and the eigenstates are trivially constructed. For $t/m=0$ the
Wannier states are localized in the center of the unit-cells, whereas for $t/m
= -1$ the Wannier states exist at the boundary of the unit-cells, see
Figs.~\ref{fig:1}b and c. Since it is not possible to continuously move Wannier
states from the position in Fig.~\ref{fig:1}b to the position in
Fig.~\ref{fig:1}c without breaking the inversion symmetry, the two cases
represent different topological phases. As it is the presence of a crystalline
symmetry that rules out a continuous transition between the two topological
phases, such topologically different atomic-limit insulators are referred to as
``symmetry-obstructed atomic insulators'' \cite{po2017,bradlyn2017}. 

One-dimensional gapped Hamiltonians with inversion symmetry, but without
particle-hole antisymmetry, have a $\ZZ$ classification
\cite{chiu2013,morimoto2013,shiozaki2014,trifunovic2017,chiu2016}, the
topological index ${\cal N}$ corresponding to the difference of the number of
occupied odd-parity bands at $k=\pi$ and $k=0$. The generator of the
classifying group is the topological equivalence class of the model
(\ref{eq:Kitaev}) with $-2t < m < 0$. There are no anomalous corner charges,
nor any other boundary boundary signatures, if ${\cal N}$ is even.

\subsection{BBH model} \label{sec:3.2}
\begin{figure}[t]%
\centering
\includegraphics*[width=.8\columnwidth]{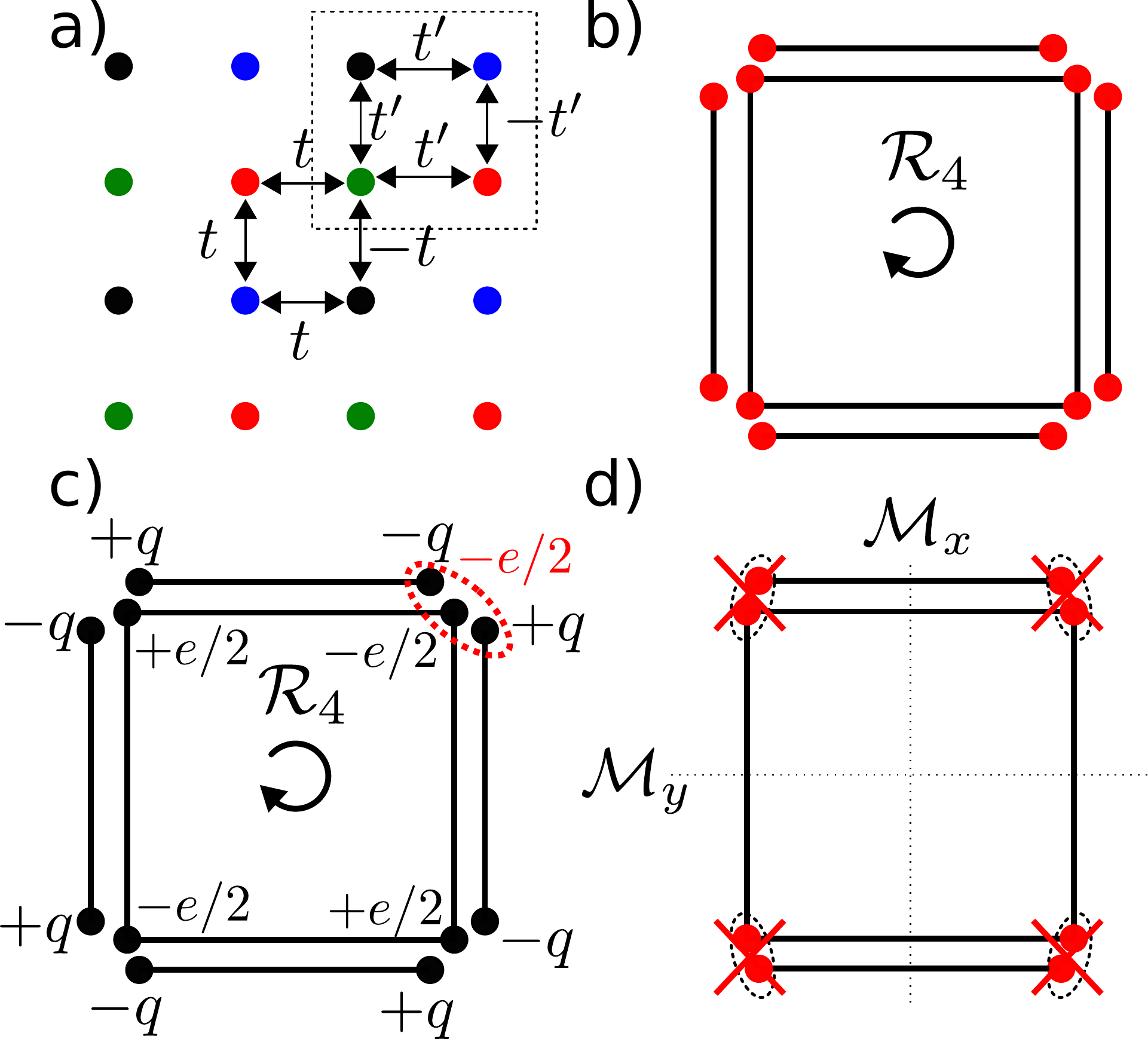}
\caption{Representation of the BBH model as a lattice model with a four-atom unit-cell and nearest-neighbor hopping (a). With fourfold rotation symmetry ${\cal R}_4$ the model exhibits zero-energy corner states in the presence of ${\cal P}$ or ${\cal C}$. These corner states are intrinsic: They cannot be removed by an ${\cal R}_4$-compatible change of boundary termination (b). With ${\cal R}_4$, but without ${\cal P}$ and ${\cal C}$, the corners may have anomalous half-integer charges. These, too, cannot be removed by an ${\cal R}_4$-compatible change of boundary termination (c). If ${\cal R}_4$ is broken, zero-energy corner states can be removed by suitably added decorations, regardless of the presence of the mirror symmetries ${\cal M}_x$ and ${\cal M}_y$ (d). }
\label{fig:BBH}
\end{figure}

The first model featuring intrinsic anomalous corner states was considered by
Benal\-cazar, Bernevig, and Hughes (BBH) \cite{benalcazar2017,benalcazar2017b}.
It is a four-band model,
\begin{align}
  \label{eq:HBBH}
  H(k_1,k_2) =&\, \Gamma_+ (t' + t \cos k_1) + \Gamma_1 t \sin k_1
   \\ \nonumber &\, \mbox{} + \Gamma_- (t' + t \cos k_2) 
  + \Gamma_2 t \sin k_2,
\end{align}
where $t$ and $t'$ are real parameters and $\Gamma_+ = \tau_1 \sigma_0$,
$\Gamma_- = -\tau_2 \sigma_2$, $\Gamma_1 = -\tau_2 \sigma_3$, $\Gamma_2 =
-\tau_2 \sigma_1$ are mutually anticommuting hermitian matrices that square to
one. The Hamiltonian (\ref{eq:HBBH}) satisfies two mirror symmetries ${\cal
M}_{x,y}$, 
\begin{align}
  H(k_1,k_2) =&\, U_{x} H(-k_1,k_2) U_{x} \nonumber \\
  =&\, U_y H(k_1,-k_2) U_y, 
\end{align}
with $U_x = \tau_1 \sigma_3$ and $U_y = \tau_1 \sigma_1$, as well as
a fourfold rotation symmetry ${\cal R}_4$ with $({\cal R}_4)^4 = -1$, 
\begin{equation}
  H(k_1,k_2) =   U_{\cal R} H(k_2,-k_1) U_{\cal R}^{\dagger},
\end{equation}
with 
\begin{equation}
  U_{\cal R} = \begin{pmatrix} 0 & \sigma_0 \\ -i \sigma_2 & 0 \end{pmatrix}.
\end{equation}
Like the previous example, the Hamiltonian (\ref{eq:HBBH}) is invariant under
time-reversal ${\cal T}$, particle-hole conjugation ${\cal P}$, and the chiral
antisymmetry ${\cal C}$,
\begin{align}
  H(k_1,k_2) =&\, U_{\cal T} H(-k_1,-k_2)^* U_{\cal T} \nonumber \\
  =&\, -U_{\cal P} H(-k_1,-k_2)^* U_{\cal P} \nonumber \\
  =&\, -U_{\cal C} H(k_1,k_2) U_{\cal C},
\end{align}
with $U_{\cal T} = 1$, $U_{\cal P} = U_{\cal C} = \tau_3$. In the presence of particle-hole symmetry, the BBH model (\ref{eq:HBBH}) may be understood as a superconductor with a two-atom unit-cell. In this case the order parameter has odd mirror parity, but it transforms trivially under rotations. Without ${\cal P}$, Eq.~(\ref{eq:HBBH}) has an intuitive explanation in terms of a tight-binding model on the square lattice with a four-atom unit-cell, see Fig.~\ref{fig:BBH}.

Equation~(\ref{eq:HBBH}) does not have the Dirac-like form of
Eq.~(\ref{eq:Kitaev}), which allows a straightforward analysis of its
topological content by counting mass terms.  However, it can be smoothly
transformed to a Hamiltonian of that form. This is most conveniently performed
in the vicinity of the gapless point at $t' = \pm t$ by parameterizing
\begin{equation}
  t' = -t + m/2
\end{equation}
and rewriting Eq.~(\ref{eq:HBBH}) as
\begin{align}
  H(k_1,k_2) =&\, \Gamma_0 \left[m + t (2 - \cos k_1 - \cos k_2)\right]
  \nonumber \\ &\, \mbox{} +
  \Gamma_1 t \sin k_1 + \Gamma_2 t \sin k_2
  \nonumber \\ &\, \mbox{} 
  - \Gamma_4 t (\cos k_1 - \cos k_2),
  \label{eq:HBBHcanonical}
\end{align}
where $\Gamma_{\pm} = -(\Gamma_0 \pm \Gamma_4)/\sqrt{2}$ and we rescaled the terms proportional to $\Gamma_{0,4}$. In the vicinity of the gapless point $m=0$, the term proportional to $\Gamma_4$ can be sent to zero without violating any of the symmetries or closing a spectral gap. The remaining three terms have the desired Dirac-like form [compare with Eq.~(\ref{eq:Kitaev})]. 

Whether the gapless point at $m=0$ represents a topological phase transition can be easily decided by inspection of the ``mass terms'' of the Hamiltonian (\ref{eq:HBBHcanonical}). There are two such mass terms: $M_1 = \Gamma_4$ and $M_2 = \tau_3 \sigma_0$. The first of these is compatible with all symmetries discussed above, except for ${\cal R}_4$, which implies that the model (\ref{eq:HBBH}) is topologically trivial unless ${\cal R}_4$ is present. The mass term $M_2$ is compatible with ${\cal T}$ and with twofold rotation symmetry ${\cal R}_2 = {\cal M}_x {\cal M}_y$ only. 

With fourfold rotation symmetry the model~(\ref{eq:HBBH}) has a nontrivial
topological phase for $-4t < m < 0$. In the presence of ${\cal P}$ or ${\cal
C}$ this phase has anomalous zero-energy (Majorana) corner states. If ${\cal
P}$ or ${\cal C}$ are broken, the corner states may be removed by a local
perturbation. In this case, the model~(\ref{eq:HBBH}) is in an obstructed
atomic-limit phase in which --- analogously to the one-dimensional example
discussed above --- anomalous half-integer corner charges
remain~\cite{benalcazar2017b,benalcazar2017,vanmiert2018}. These corner states
or corner charges are {\em intrinsic}: they are a consequence of the topology
of the bulk band structure and they exist independently of the lattice
termination, as long as the termination is compatible with the fourfold
rotation symmetry ${\cal R}_4$~\cite{ono2019d}. To see this, observe that the
relevant changes of boundary termination correspond to ``decorating'' the
boundaries with one-dimensional chains with a gapped excitation spectrum.
Although such ``decorations'' may have zero-energy (Majorana) end states (with
${\cal P}$ or ${\cal C}$) or fractional end charges (without ${\cal P}$ and
${\cal C}$), the requirement that the decoration be ${\cal R}_4$-compatible
means that the net number of zero-energy states added to each corner is even
(with ${\cal P}$ or ${\cal C}$) or that the net charge added to each corner is
an integer (without ${\cal P}$ or ${\cal
C}$)~\cite{vanmiert2018,benalcazar2019}, see Fig.~\ref{fig:BBH}. 

The bulk band structure of Eq.~(\ref{eq:HBBH}) is always trivial if ${\cal
R}_4$ is broken.\footnote{References \cite{benalcazar2017b,benalcazar2017} show
that the model (\ref{eq:HBBH}) exhibits a phase with a quantized quadrupole
moment in the presence of the two mirror symmetries ${\cal M}_x$ and ${\cal
M}_y$. The quantized quadrupole moment is associated with the bulk band
structure. Despite it being quantized for a generic gapped crystalline phase
protected by ${\cal M}_x$ and ${\cal M}_y$, the quadrupole moment is, however,
not a topological invariant, since it can change its value without a closing of
the bulk excitation gap.} Nevertheless, if ${\cal R}_4$ is broken, zero-energy
corner states may also exist if ${\cal P}$ or ${\cal C}$ is present. In this
case the corner states are {\em extrinsic}: Their existence depends on the
lattice termination and they may be removed by decorating the boundaries with
one-dimensio\-nal gapped chains, see Fig.~\ref{fig:BBH}. The same applies to
the existence of half-integer corner charges for the case that ${\cal P}$ and
${\cal C}$ are broken. The presence or absence of the two mirror symmetries
${\cal M}_{x,y}$ does not affect these conclusions.

Two-dimensional gapped Hamiltonians with intrinsic anomalous corner sta\-tes
realize a {\em second-order} topological phase. The example of the BBH model
shows that the identification of the relevant non-spatial and crystalline
symmetries is key to deciding whether or not a given band structure represents
a second-order topological phase. Indeed, the very same model (\ref{eq:HBBH})
with the same choice of parameters $-4 t < m < 0$ may be a second-order phase,
an obstructed atomic-limit phase, or a trivial phase depending on whether or
not ${\cal P}$ or ${\cal C}$ and fourfold rotation symmetry ${\cal R}_4$ are
enforced.

As in the one-dimensional example discussed in the previous Subsection, the BBH
model is part of a larger family of topological Hamiltonians. For example,
two-dimensional Hamiltonians with ${\cal P}$ and ${\cal R}_4$ have a $\ZZ^3$
classification \cite{teo2013,benalcazar2014,cornfeld2019}, where one factor
$\ZZ$ describes a first-order topological phase with chiral (Majorana) edge
mo\-des. The second factor $\ZZ$ describes a sequence of topological phases for
which the topological class of the BBH model (\ref{eq:HBBH}) with $-4t < m < 0$
is the generator. There are no anomalous corner states if the corresponding
topological index is even, consistent with the $\ZZ_2$ nature of the Majorana
corner modes.  The last factor $\ZZ$ describes additional atomic-limit phases
without boundary signatures.

\subsection{Three-dimensional example}
As a third example we discuss a three-dimensional generalization of the BBH
model originally proposed by Schindler {\em et al.} \cite{schindler2018},
\begin{align}
  H(\vk) =&\, \Gamma_0 \left[m + t (3 - \cos k_1 - \cos k_2 - \cos k_3)\right]
  \nonumber \\ &\, \mbox{} +
  \Gamma_1 t \sin k_1 + \Gamma_2 t \sin k_2 + \Gamma_3 t \sin k_3
  \nonumber \\ &\, \mbox{} 
  + \Gamma_4 t' (\cos k_1 - \cos k_2),
  \label{eq:HBBHcanonical3}
\end{align}
where $\Gamma_0 = \tau_3 \sigma_0$, $\Gamma_k = \tau_1 \sigma_k$, $k=1,2,3$, and $\Gamma_4 = \tau_2 \sigma_0$. As in the previous example, see Eq.~(\ref{eq:HBBHcanonical}), the last term proportional to $\Gamma_4$ may be omitted without closing the spectral gap or violating any of the symmetries. Without this last term, the model (\ref{eq:HBBHcanonical3}) is invariant under time-reversal ${\cal T}$, 
\begin{equation}
  H(\vk) = U_{\cal T} H(-\vk)^* U_{\cal T},
\end{equation}
with $U_{\cal T} = \tau_0 \sigma_2$, as well as a four-fold rotation ${\cal R}_4$,
\begin{equation}
  H(k_1,k_2,k_3) = U_{\cal R} H(k_2,-k_1,k_3) U_{\cal R}^{-1},
\end{equation}
with $U_{\cal R} = \tau_0 e^{i \pi \sigma_3/4}$. The model
(\ref{eq:HBBHcanonical3}) has a single mass term $M = \Gamma_4$, which is
antisymmetric under ${\cal T}$ and symmetric under ${\cal R}_4$. 
\begin{figure}[t]%
	\begin{center}
		\includegraphics*[width=.6\columnwidth]{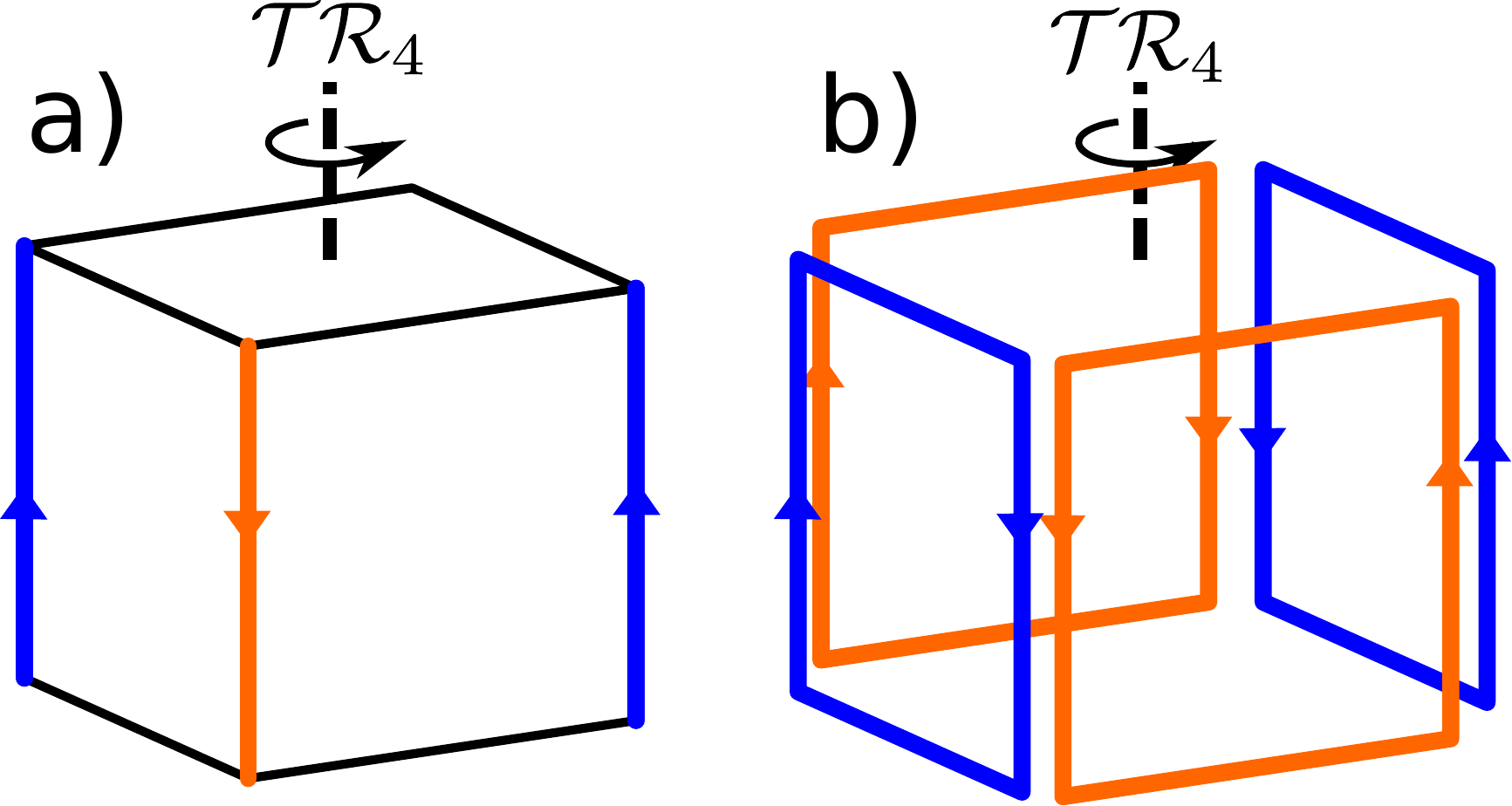}
		\caption{Configuration of chiral hinge modes of three-dimensional crystal with ${\cal TR}_4$ symmetry (a). Configurations of chiral hinge modes that can be obtained by ``decorating'' the boundary have an even number of chiral modes at each hinge (b). } \label{fig:2}
	\end{center}
\end{figure}

In the presence of ${\cal T}$ the gapless point $m=0$ is a topological phase
transition between phases with and without gapless surface states, irrespective
of the fourfold rotation symmetry ${\cal R}_4$.  Schindler {\em et al.}
observed that the mass term $M = \Gamma_4$ is not only antisymmetric under
${\cal T}$, but also under the product ${\cal TR}_4$, so that if ${\cal TR}_4$
is a good symmetry, the gapless point $m=0$ still separates topologically
different band structures, even if ${\cal T}$ and ${\cal R}_4$ symmetries are
broken individually. In this case, the topological phase is not characterized
by gapless surface states (because these are gapped out on a generic surface if
time-reversal symmetry is broken), but by a chiral gapless mode running along
the crystal ``hinges'', see Fig.~\ref{fig:2}. This gapless mode is a
manifestation of the topological nature of the bulk band structure and it can
not be removed by changing the crystal termination. This can be understood by
noticing that the relevant change of surface termination corresponds to
decorating each of the four crystal faces by a two-dimensional quantized Hall
insulator, which will change the number of chiral modes running along a hinge
by an even number if the surface decoration is compatible with the ${\cal
TR}_4$ symmetry, see Fig.~\ref{fig:2}b. A boundary pattern shown in
Fig.~\ref{fig:2}a, which has an odd number of chiral modes at each hinge, is
``anomalous'' --- it cannot exist without a topologically nontrivial
three-dimensional bulk. Because of the existence of anomalous boundary states
of codimension two, the model (\ref{eq:HBBHcanonical3}) is a {\em second-order}
topological phase.

\begin{figure}[t]%
	\begin{center}
		\includegraphics*[width=.6\columnwidth]{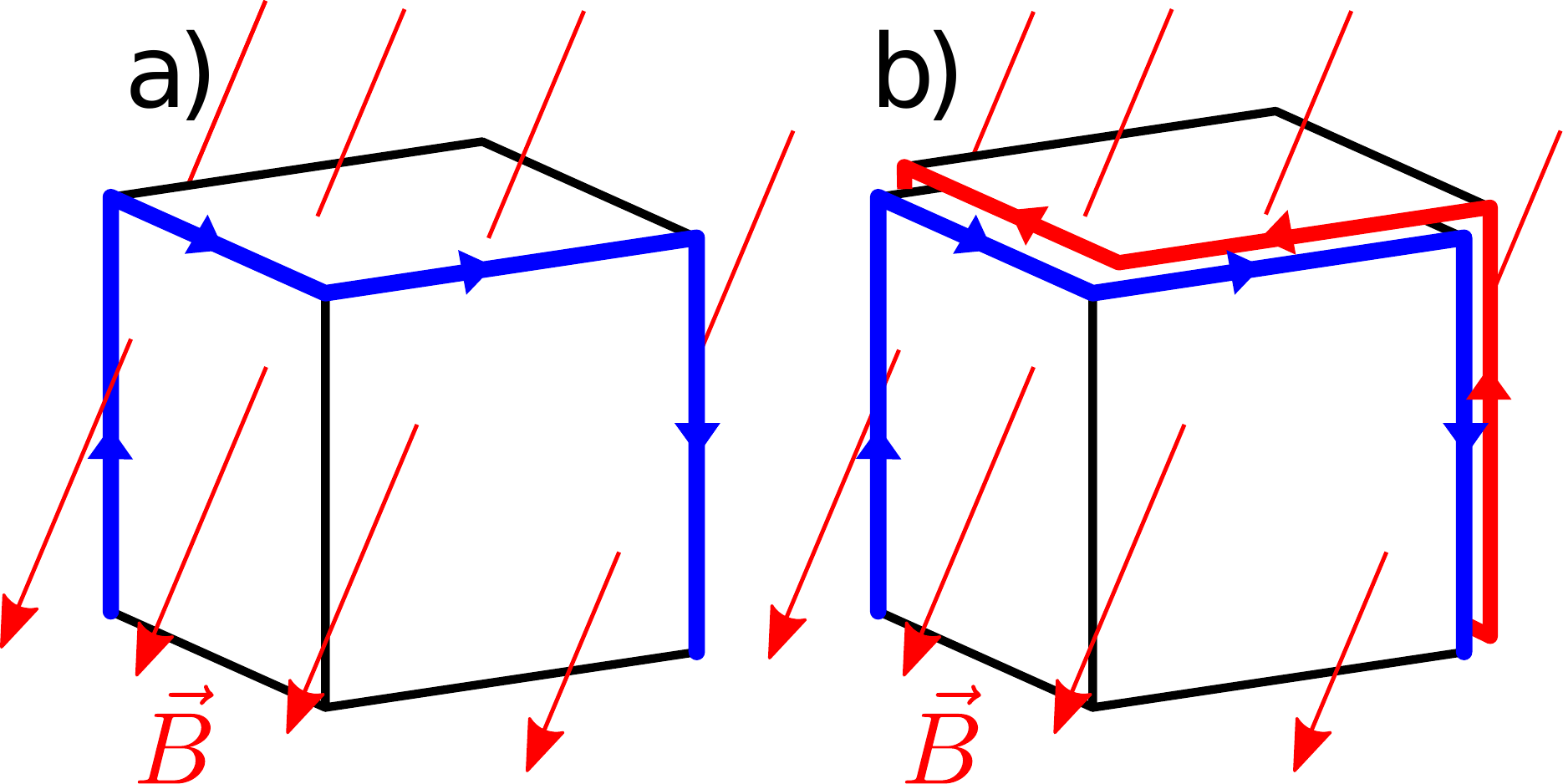}
		\caption{Schematic picture of an ``extrinsic'' second-order topological
			insulator consisting of a three-dimensional topological insulator
			placed in a magnetic field in a generic direction, as proposed by Sitte
			\textit{et al.}~\cite{sitte2012} (a). Each surface has a finite flux
			and there are chiral modes along hinges that touch two faces with
			opposite sign of the magnetic flux. The gapless hinge modes may be
			removed by decorating some of the crystal faces with a
		two-dimensional anomalous quantized Hall insulator (b). }
		\label{fig:3}
	\end{center}
\end{figure}

The same boundary phenomenology can exist in the absence of crystalline
symmetries. An early example was proposed by Sitte {\em et al.}, who showed
that chiral hinge modes are generic for a topological insulator in an external
magnetic field~\cite{sitte2012}. Such hinge modes are not anomalous, however,
since they can be removed by an appropriate surface decoration, see Fig.\
\ref{fig:3}b. For this reason, the hinge states of the ${\cal TR}_4$-symmetric
model of Eq.~(\ref{eq:HBBHcanonical3}) are {\em intrinsic}, whereas hinge
states that appear in the absence of a crystalline symmetry are {\em
extrinsic}. 

Without the term proportional to $\Gamma_4$, the
model~(\ref{eq:HBBHcanonical3}) is not only invariant under time reversal and
fourfold rotation, but it also satisfies three anticommuting mirror symmetries
${\cal M}_i:\, k_i \to -k_i$, with $U_{ {\cal M}_i} = \Gamma_4 \Gamma_i$, $i=1,2,3$, and
inversion symmetry ${\cal I}: \vk \to -\vk$, with $U_{\cal I} = \Gamma_0$.
Since the mass term $\Gamma_4$ is incompatible with each of these crystalline
symmetries, imposing mirror or inversion symmetry while breaking ${\cal T}$
also results in a second-order topological phase for $-6t < m <
0$~\cite{schindler2018,langbehn2017,geier2018,khalaf2018,khalaf2018b,trifunovic2019}.

\subsection{Higher-order topological phases}

In general, a crystalline topological phase is called an $n$th-order
topological phase if it has anomalous boundary signatures of codimension $n$ if
the boundary as a whole respects the crystalline symmetries. With this
definition, a topological phase that does not rely on crystalline symmetries
for its protection is a first-order topological phase. Since topological phases
without crystalline symmetries always have a first-order boundary signature,
higher-order topological phases are necessarily crystalline topological phases.
The above definition of the order of a topological crystalline phase assumes a
crystal termination for which the codimension of the boundary states is
maximal. For special choices of the termination, an $n$-th order topological
phase may have boundary signatures of codimension smaller than $n$. For
example, a crystal with a face invariant under the point-group symmetry $G$
hosts codimension-one boundary states if the bulk band structure has a
nontrivial topology.  For a higher-order topological phase these
codimension-one boundary states disappear for a generic orientation of the
faces, leaving boundary states of a larger codimension behind, see
Fig.~\ref{fig:4}.  Additionally, as the example in Fig.~\ref{fig:5}a-b
shows, even with an identical orientation of the faces, a given crystal can
exhibit boundary states of different codimension depending on the details
of the termination. Whereas for the former example, the convention of
assigning the order of a phase according to the maximal codimension
follows naturally from the observation that a termination without symmetry
invariant faces is more generic than the one with such faces, for the
example in Fig.~\ref{fig:5}a-b determining the order of the topological
phase according to the maximal codimension of the anomalous boundary
states is a matter of convention. 
Lastly, we note that in the literature, topological phases with anomalous
corner \textit{charges} (as opposed to corner \textit{states}) are sometimes
also referred to as higher-order phases, although such obstructed atomic-limit
phases fall outside the boundary-based classification scheme we will use in
this review. (They are captured in the $K$-theory-based classification scheme
of the bulk band structure.) 

\subsection{Experimental realizations of higher-order band structures}\label{sec:exp}
The existence of corner modes in the BBH model was demonstrated experimentally
in various classical systems. These include electrical
\cite{imhof2018,serra-garcia2019} and microwave \cite{peterson2018} circuits,
as well as coupled mechanical oscillators~\cite{serra-garcia2018}. Although the
spatial symmetries and the chiral antisymmetry required to pin the frequency of
the corner modes to the center of an excitation gap are not natural symmetries
for these platforms, the near-complete control over device parameters ensures
that these symmetries can be implemented experimentally to a sufficiently high
degree. Corner modes of true quantum-mechanical origin were observed in an
artificial electronic lattice obtained by placing CO molecules on a Cu(111)
surface~\cite{kempkes2019}, although the specific lattice model implemented in
Ref.~\onlinecite{kempkes2019} has a topologically trivial band structure with
accidental non-anomalous corner modes.

\begin{figure}[t]%
	\begin{center}
		\includegraphics*[width=\columnwidth]{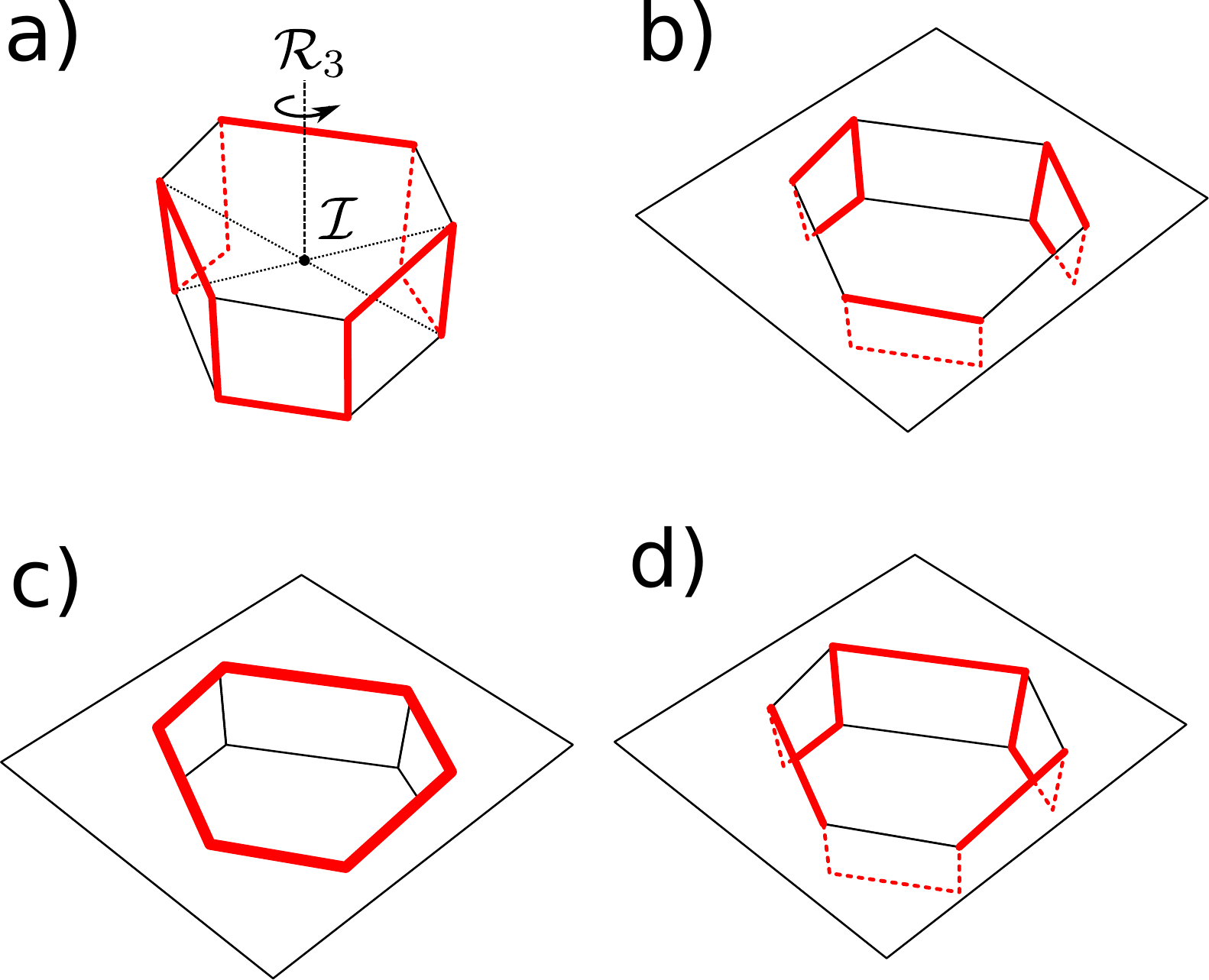}
		\caption{a): Hexagonal Bi pillar with threefold rotation axis, inversion center, and arrangement of anomalous hinge modes characteristic of a second-order band struture \cite{Schindler2018b}. b): Hexagonal terrace with the edge states consistent with (a). c) and d): Anomalous one-dimensional modes at the edges of a one-atomic layer high terrace on the surface of a weak topological insulator. Whether the one-dimensional modes are centered at the upper or lower end of the terrace is not fixed by the weak topology alone and may depend on details of the termination or on the combined presence of weak and second-order topology.}
		\label{fig:bismuth}
	\end{center}
\end{figure}

To date, experimental evidence of higher-order topology of three-dimensional
band structures was reported for two materials only: Elemental
Bismuth~\cite{Schindler2018b,nayak2019} and the Van der Waals material
Bi$_4$Br$_4$~\cite{noguchi2020}. In Bi$_4$Br$_4$, which is a two-dimensional
stacking of quasi one-dimensional molecules, helical hinge states are protected
by a twofold rotation symmetry along the molecular axis. Noguchi {\em et al.}
demonstrate the existence of one-dimensional hinge states in Bi$_4$Br$_4$ using
angle-resolved photoemission spectroscopy~\cite{noguchi2020}, a technical
tour-de-force given the low dimensionality of the feature to be resolved
spectroscopically. Hinge states in Bi are protected by inversion symmetry
${\cal I}$. Bismuth has an additional threefold rotation symmetry ${\cal R}_3$,
leading to a characteristic hexagonal pattern of helical hinge states for a Bi
flake cut perpendicular to the rotation axis, see Fig.~\ref{fig:bismuth}a. The
experimental demonstration that Bi has a higher-order band structure is
complicated by the fact that Bi is a semimetal with indirect band overlap.
Evidence that Bi has a higher-order band structure is based on a measurement of
the current-phase relationship for the Josephson current through Bi pillars
\cite{murani2017,Schindler2018b}, which has a sawtooth-like contribution
characteristic of ballistic one-dimensional channels. The observation of
one-dimensional states at three out of six step edges of a hexogonal terrace on
a Bi surface provides further evidence for the higher-order topology of the Bi
band structure. This experimental signature is not unambiguous, however.
Theoretically, for a terrace (as opposed to a free-standing flake) one expects
the pattern of edge modes shown in Fig.~\ref{fig:bismuth}b if the band
structure has second-order topology. This is most easily seen in the
domain-wall picture (see next Subsection) by observing that each crystal face
has a well-defined sign of its unique surface mass term that depends on the
orientation of that surface only.~\footnote{Note that upon placing the ﬂake of
Fig~\ref{fig:bismuth}a into the terrace of Fig.~\ref{fig:bismuth}b it appears
as if an odd number of hings modes is left behind, as in
Fig.~\ref{fig:bismuth}c. This is not the case. Inclusion of the hopping between
the ﬂake and the terrace closes and reopens the surface gaps at the boundary
between the flake and the terrace, which removes these hinge states. This is
analogous to the situation that occurs when placing the two copies of the ﬂake
from Fig.~\ref{fig:bismuth}a on top of each other.} The presence of an even
number of modes at each edge of the terrace means that the edge modes no longer
have topological protection if the terrace is only one or a few atomic layers
thick.  Based on scanning tunneling spectroscopy experiments close to a screw
dislocation, Ref.~\onlinecite{nayak2019} finds that Bi has a nontrivial weak
topology. With weak topology, which may or may not occur in combination with
second-order topology of the bulk band structure, a one atomic-layer high step
edge hosts an odd number of helical modes. The mode patterns corresponding to
weak topology, such as those of Fig.~\ref{fig:bismuth}c and d, may be difficult
to distinguish experimentally from the mode pattern of Fig.~\ref{fig:bismuth}b.

\subsection{Domain-wall picture}
\label{sec:DW}
\begin{figure}[t]%
	\begin{center}
		\includegraphics*[width=.8\columnwidth]{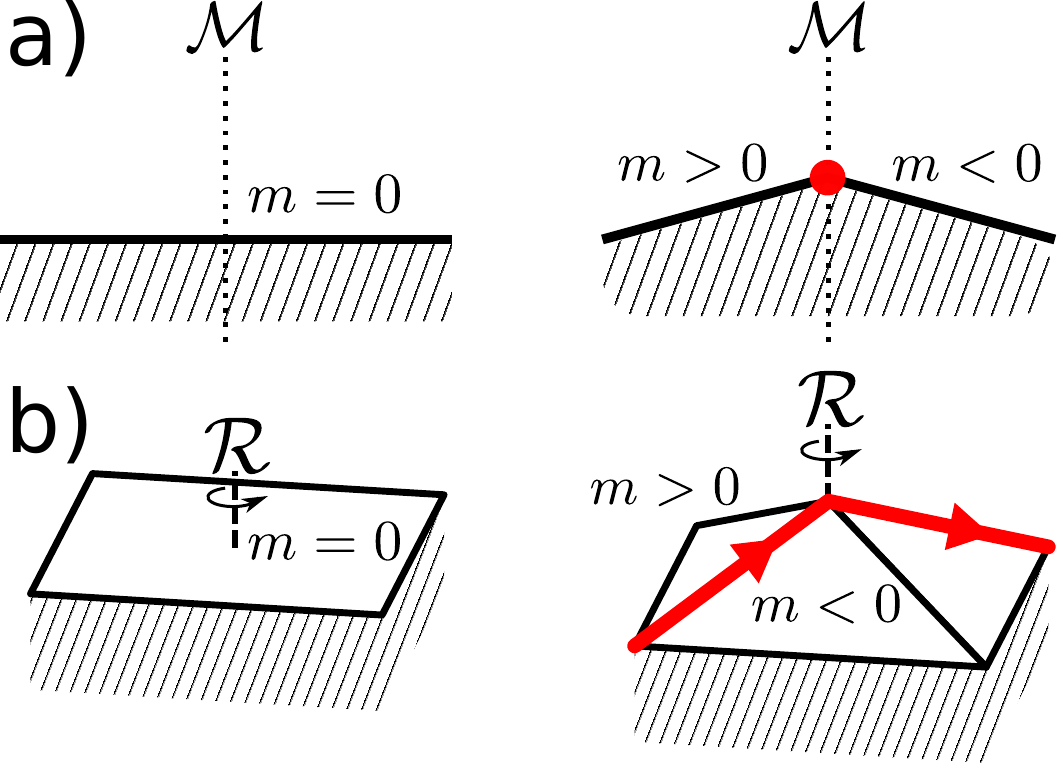}
		\caption{(a) A two-dimensional crystal with a mirror-symmetric edge, at which the mirror symmetry acts as a local symmetry (left) and a generic mirror-symmetric termination, at which the mirror symmetry acts as a ``global'' symmetry only (right). (b) A three-dimensional crystal with a rotation-invariant surface, at which the rotation symmetry acts as a local symmetry (left) and a generic fourfold rotation-symmetric termination, at which the rotation symmetry acts as a global symmetry (right). In both panels, a configuration of boundary mass terms is shown that corresponds to a second-order phase. }
	\label{fig:4}
	\end{center}
\end{figure}

To understand why a topological band structure with crystalline symmetries can
give rise to higher-order boundary states,
Refs.~\onlinecite{langbehn2017,geier2018,khalaf2018,khalaf2018b,trifunovic2019}
propose a ``domain-wall picture''. In its original form, the domain-wall
picture applies if the crystalline symmetry group $G$ admits a boundary
orientation that is invariant under $G$.  This is the case, {\em e.g.}, for a
mirror symmetry in two dimensions or for a rotation symmetry in three
dimensions, see Fig.~\ref{fig:4}. For such an invariant boundary, the
crystalline symmetry continues to act as a {\em local} symmetry at the
boundary. Hence, the bulk-boundary correspondence for non-spatial symmetries is
applicable and guarantees the existence of an anomalous boundary state on that
crystal face. The low-energy theory of that anomalous $(d-1)$-dimensional
boundary state has the form of a Dirac Hamiltonian,
\begin{equation}
  H_{\partial}(k_1,\ldots,k_{d-1}) = \sum_{i=1}^{d-1} \gamma_i k_i,
\end{equation}
with anticommuting gamma matrices $\gamma_i^2=1$, $i=1,\ldots,d-1$. Again, we search for anticommuting mass terms $\mu_j^2=1$ with $\{\mu_j,\gamma_i\} = 0$, $i=1,\ldots,d-1$, $j=1,\ldots,n$. If the bulk topological phase is not a first-order phase there must be at least one such mass term. The mass terms transform under a real representation $O_n$ of the crystalline symmetry group $G$, which cannot be the trivial representation, since otherwise $H_{\partial}$ can be gapped out by a symmetry-preserving perturbation. 

Higher-order boundary states appear when we consider deformations of the invariant boundary, so that $G$ no longer acts locally, but continues to act on the crystal boundary as a whole. Examples of such deformations are shown in Fig.~\ref{fig:4}. Given the number $n$ of mass terms and the representation $O_n$ of $G$ one can derive the pattern of anomalous higher-order boundary states. Such boundary states of are of second order if $d=2$. They are also of second order if $d=3$ and $n=1$ or if $G$ leaves a one-dimensional subset of the deformed crystal face invariant (as is the case for, {\em e.g.}, a mirror symmetry). Otherwise the boundary states are of third order in three dimensions.

A ``trick'' to extend this argument to symmetry groups $G$ without invariant
boundary orientation, such as inversion symmetry, was proposed in
Ref.~\onlinecite{geier2018}. The trick involves considering a $(d+1)$-dimensional
topological crystalline band structure with a symmetry group $G'$ obtained by
acting with $G$ on the first $d$ coordinates, while leaving the
$(d+1)$th coordinate unchanged. For the
$(d+1)$-dimensional crystal the conditions of the domain-wall
argument are obviously fulfilled, so that one can establish the existence of
higher-order boundary states using the domain-wall picture outlined above.
Reference~\onlinecite{geier2018} then makes use of an isomorphism between
$(d+1)$-dimensional topological band structures with symmetry group $G'$ and
$d$-dimensional topological band structures with symmetry group $G$ that was
originally derived for non-crystalline topological phases by Fulga
{\em et al.}~\cite{fulga2011}. Since this isomorphism preserves the order of
the anomalous boundary states~\cite{trifunovic2017,geier2018}, one can directly
infer the existence of higher-order boundary states for the $d$-dimensional
crystal with symmetry group $G$.

As an example, we consider the crystal described by
Eq.~(\ref{eq:HBBHcanonical3}). The ${\cal TR}_4$ symmetry leaves surfaces at
constant $z$ invariant. The low-energy surface Hamiltonian has the form
\begin{equation}
  H_{\partial}(k_1,k_2) = k_1 \gamma_1 + k_2 \gamma_2,
\end{equation}
with $\gamma_1 = \sigma_1$, $\gamma_2 = \sigma_3$. The surface Hamiltonian $H_{\partial}$ satisfies the product ${\cal TR}_4$ of time-reversal and fourfold rotation symmetry,
\begin{equation}
  H_{\partial}(k_1,k_2) = U_{\cal TR} H_{\partial}(-k_2,k_1)^* U_{\cal TR}^{\dagger},
\end{equation}
with $U_{\cal TR} = e^{-i \pi \sigma_2/4}$. There is a single mass term $\mu = \sigma_2$, which changes sign under ${\cal TR}_4$. If one then deforms the invariant surface as in Fig.~\ref{fig:4}b, the faces related by fourfold rotation have opposite masses, so that there are domain walls with a sign change of the mass term at ``hinges'' between these faces \cite{schindler2018}. The gapless chiral modes run along the domain walls. The same argument can be used if the model (\ref{eq:HBBHcanonical3}) is considered with a mirror symmetry ${\cal M}_x$ or ${\cal M}_y$ instead of with ${\cal TR}_4$ symmetry \cite{schindler2018,langbehn2017}.

\subsection{Boundary states from Dirac-like bulk Hamiltonian} \label{sec:boundary_dirac}
The low-energy Dirac theory of the boundary and the transformation behavior of the boundary mass terms under the crystalline symmetry group can be obtained by direct calculation from the Dirac-like form of the bulk band structure. The starting point is the $2b$-band Dirac Hamiltonian 
\begin{align}
  H(\vk)&= m\Gamma_0+\sum_{j=1}^{d} k_j \Gamma_j,
	\label{eq:HdG}
\end{align}
which is, {\em e.g.}, the low-energy limit of the models (\ref{eq:Kitaev}), (\ref{eq:HBBHcanonical}), or (\ref{eq:HBBHcanonical3}). Here the matrices $\Gamma_j$ are mutually anticommuting $2b \times 2b$ matrices that satisfy $\Gamma_j^2 = 1$, $j=0,1,\ldots,d$.

The crystal boundary is modeled as the interface between regions with negative and positive $m$, with $m$ negative in the interior of the crystal. Near the sample boundary, the Hamiltonian (\ref{eq:HdG}) has the form
\begin{equation}
  H = m(x_{\perp}) \Gamma_0 -i \hbar \vGamma \cdot \partial_{\vr},
  \label{eq:H0boundary}
\end{equation}
where $x_{\perp} = \vn \cdot \vr$ is the coordinate transverse to the boundary, $\vn$ is the outward-pointing normal, and $\vGamma$ is a $d$-compo\-nent vector containing the matrices $\Gamma_j$, $j=1,\ldots,d$. We choose $m(x_{\perp}) > 0$ for $x_{\perp} > 0$ and $m(x_{\perp}) < 0$ for $x_{\perp} < 0$, so that the sample interior corresponds to negative $x_{\perp}$. The Hamiltonian (\ref{eq:H0boundary}) admits $b$ gapless boundary modes, the projection operator to the space of allowed $2b$-component spinors being
\begin{equation}
  P(\vn) = \frac{1}{2}[i (\vn \cdot \vGamma) \Gamma_0 + 1].
  \label{eq:Pn}
\end{equation}

The effective $b$-band low-energy surface Hamiltonian is obtained using the
projection operator $P(\vn)$. To illustrate this procedure, we consider a
hypothetical circular or spherical ``crystal'' of radius $R$ and use the polar
coordinate $\phi$ or spherical coordinates $(\theta,\varphi)$ to parameterize
$\vn$ and the crystal boundary for $d=2$ or $d=3$,
respectively~\cite{khalaf2018}. We write the projection operator (\ref{eq:Pn})
as
\begin{align}
  P(\vn) =&\, V_d(\vn) P_d V_d(\vn)^{-1}, & d=2,3,
\end{align}
with $P_2 = P(\ve_x)$, $P_3 = P(\ve_z)$, $V_2(\vn) = e^{\phi \Gamma_2
\Gamma_1/2}$, and $V_3(\vn) = e^{(\pi-\theta) \Gamma_1 \Gamma_3/2} e^{\varphi
\Gamma_2 \Gamma_1/2}$.  The projected Hamiltonian $H$ at the boundary then
reads, 
\begin{align*}
  P(\vn) H P(\vn) =&\,
  V_2(\vn) P_2  \left( -i \Gamma_2 \frac{1}{R} \frac{\partial}{\partial \phi} \right) P_2 V_2(\vn)^{-1}
\end{align*}
if $d=2$ and
\begin{align*}
  P(\vn) H P(\vn) =&\,
  V_3(\vn) P_3 
  \nonumber \\ &\, \mbox{} \times
  \left(-i \Gamma_1 \frac{1}{R} \frac{\partial}{\partial \theta}
  -i \Gamma_2 \frac{1}{R \sin \theta} \frac{\partial}{\partial \varphi}
   \right)
  \nonumber \\ &\, \mbox{} \times
  P_3 V_3(\vn)^{-1}
\end{align*}
if $d=3$. Mass terms $M_i$, $i=1,\ldots,n$, of the bulk band structure
(\ref{eq:HdG}) may be added as a position-dependent perturbation to the
surface. Although such mass terms locally violate the crystalline symmetry
group $G$, the position dependence of the surface perturbation ensures
compatibility with $G$ for the crystal as a whole. The matrices $\gamma_2 =
P_2 \Gamma_2 P_2$ (for $d=2$) or $\gamma_{1,2} = P_3 \Gamma_{1,2} P_3$ (for
$d=3$), combined with $\mu_j = P_d M_j P_d$, $j=1,\ldots,n$, form a set of
anticommuting gamma matrices and mass terms of effective dimension $b$. This
gives the effective boundary Hamiltonian
\begin{align} \label{eq:Hpartial}
  H_{\partial} =&\, \gamma_2 \left( -i \frac{1}{R} \frac{\partial}{\partial \phi} \right) + \sum_{j=1}^{n} m_j(\phi) \mu_j, & d=2, \nonumber \\
  H_{\partial} =&\, 
  \gamma_1 \left(-i \frac{1}{R} \frac{\partial}{\partial \theta} \right)
  + \gamma_2 \left( -i \frac{1}{R \sin \theta} \frac{\partial}{\partial \varphi}
   \right)
  \\ &\, \nonumber  \mbox{} 
  + \sum_{j=1}^{n} m_j(\theta,\varphi) \mu_j, & d=3.
\end{align}
The position dependence of the prefactors $m_j$ must be chosen such that the
boundary Hamiltonian $H_{\partial}$ as a whole is compatible with the
crystalline symmetry group $G$.  Hereto, we note that the induced
representation $u_{g}$ of the crystalline symmetry operation $g$ on the
boundary Hamiltonian gives an $n$-dimensional real representation $O_{n}$ of
$G$,
\begin{equation}
  u_g \mu_j u_{g}^{-1} = \sum_{j'=1}^{n} (O_n(g))_{jj'} \mu_{j'},
\end{equation}
so that compatibility with the crystalline symmetry group is ensured if the
functions $m_j$ satisfy the requirements
\begin{equation}
  m_j(\vn) = \sum_{j'=1}^{n} m_{j'}(g \vn) O_n(g)_{j'j}.
\end{equation}

As a first example, we illustrate this procedure for the BBH
model~(\ref{eq:HBBH}). If the BBH model is written in the
form~(\ref{eq:HBBHcanonical}), a low-energy Dirac Hamiltonian of the
form~(\ref{eq:HdG}) is immediately obtained. There is only one mass term $M =
\Gamma_4$ that is invariant under particle-hole conjugation ${\cal P}$ or the
chiral antisymmetry ${\cal C}$. This mass term changes sign under the fourfold
rotation operation ${\cal R}_4$. It follows that the boundary Hamiltonian
$H_{\partial}$ is of the form (\ref{eq:Hpartial}) with the condition
\begin{equation}
  m(\phi+\pi/2) = -m(\phi) \label{eq:mcond}
\end{equation}
to ensure compatibility with respect to ${\cal R}_4$. The condition
(\ref{eq:mcond}) implies the existence of four ``domain walls'' at which
$m(\phi)$ changes sign. These domain walls each host an anomalous zero-energy
state.
\begin{figure}[t]%
	\begin{center}
		\includegraphics*[width=0.9\columnwidth]{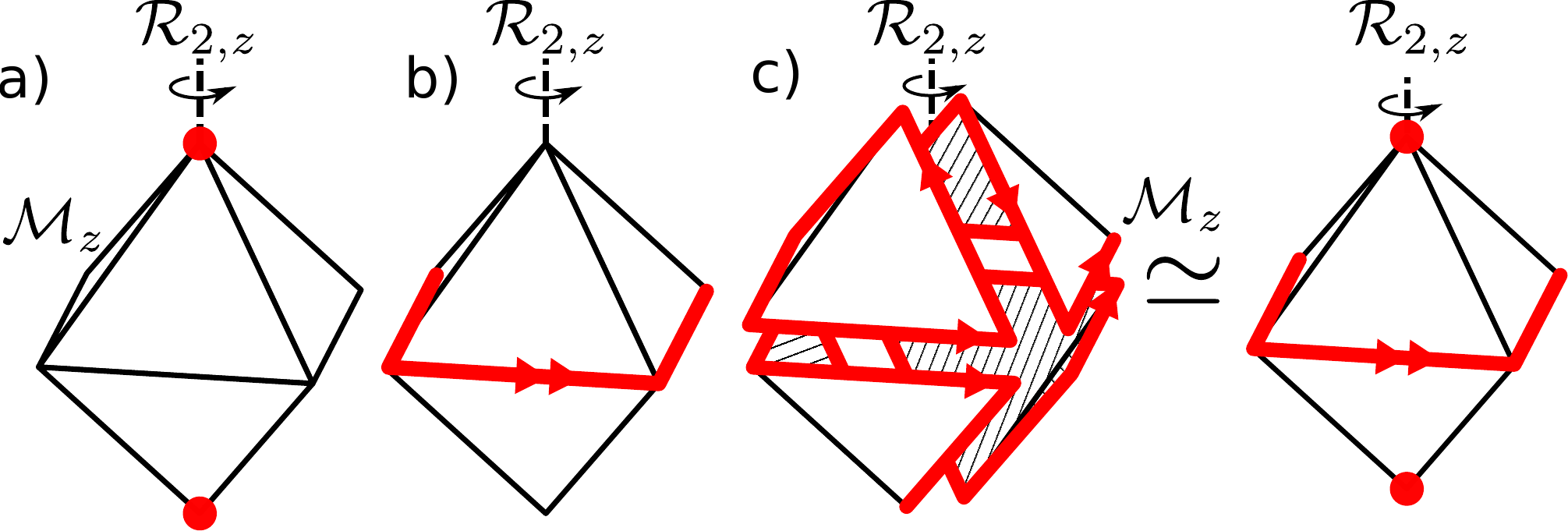}
		\caption{A topological three-dimensional superconductor with mirror and twofold rotation symmetry admits Majorana corner modes at the rotation axis (a) or a pair of co-propagating Majorana hinge modes at the mirror plane (b). The difference between these two boundary signatures is a matter of crystal termination: The higher-order boundary modes in a crystal that has corner states as well as hinge modes are extrinsic (c). }
		\label{fig:5}
	\end{center}
\end{figure}

The second example (which will be considered again in Sec.~\ref{sec:5})
is a three-dimensio\-nal odd-parity superconductor with point-group $C_{2h}$,
which is generated by commuting twofold rotation and mirror symmetries 
${\cal R}_{2,z}$ and ${\cal M}_z$. The rotation symmetry is around the 
$z$ axis; the mirror reflection is in the $xy$ plane, see Fig.~\ref{fig:5}.
Using the convention that both crystalline symmetries square to one, ${\cal
R}_{2,z}$/${\cal M}_z$ commute/anticommute with particle-hole conjugation
${\cal P}$, respectively.
This model is described by an eight-band Dirac Hamiltonian of the form
(\ref{eq:HdG}) with $\Gamma_0 = \tau_2 \sigma_0 \rho_0$, $\Gamma_1 = \tau_1
\sigma_3 \rho_0$, $\Gamma_2 = \tau_1 \sigma_1 \rho_0$, and $\Gamma_3 = \tau_3
\sigma_0 \rho_0$. The relevant symmetries are represented by $U_{\cal P} = 1$,
$U_{\cal M} = \tau_2 \sigma_2 \rho_2$, and $U_{\cal R} = \tau_0 \sigma_2
\rho_2$. There are two ${\cal P}$-symmetric mass terms that break the
crystalline symmetry: $M_1 = \tau_1 \sigma_2 \rho_1$ and $M_2 = \tau_1 \sigma_2
\rho_3$. Both mass terms are symmetric under ${\cal M}_z$ but antisymmetric
under ${\cal R}_{2,z}$. Hence, the effective surface theory is of the form
(\ref{eq:Hpartial}) with the condition

\begin{equation}
  m_j(\theta,\varphi) = -m_j(\theta,\varphi+\pi),\ \ j=1,2.
\end{equation}
Such a mass term has a singular ``vortex''-like structure at the poles at $\theta=0$, $\pi$, resulting in the presence of protected zero modes there, see Fig.~\ref{fig:5}a. There are no protected second-order boundary states, since generically at least one of the two mass terms is nonzero away from the poles.

It is interesting to point out that in this example one could also have chosen the single mass term $M = \tau_1 \sigma_2 \rho_0$. This mass term does not admit any further anticommuting mass terms. Since $M$ is odd under ${\cal M}_z$ but even under ${\cal R}_{2,z}$, one would then have concluded that this model has a pair of co-propagating chiral Majorana modes at the mirror plane, see Fig.~\ref{fig:5}b. The difference between this boundary signature and the third-order boundary signature with Majorana zero modes at the twofold rotation axis is extrinsic, since it corresponds to a trivial bulk band structure. Indeed, by explicit construction, one verifies that it is possible to construct a boundary decoration that has Majorana corner modes at the rotation axis and two co-propagating chiral modes at the mirror plane, see Fig.~\ref{fig:5}c. Addition of this boundary decoration switches between the boundary signatures of Figs.~\ref{fig:5}a and b. Since the codimension-$2$ boundary signature of Fig.~\ref{fig:5}b can be eliminated in favor of the codimension-$3$ boundary of Fig.~\ref{fig:5}a by a suitable choice of termination, this model must be considered a {\em third-order} topological band structure.

\subsection{Boundary-resolved classification}
\begin{figure}[t]%
	\begin{center}
		\includegraphics[width=0.95\columnwidth]{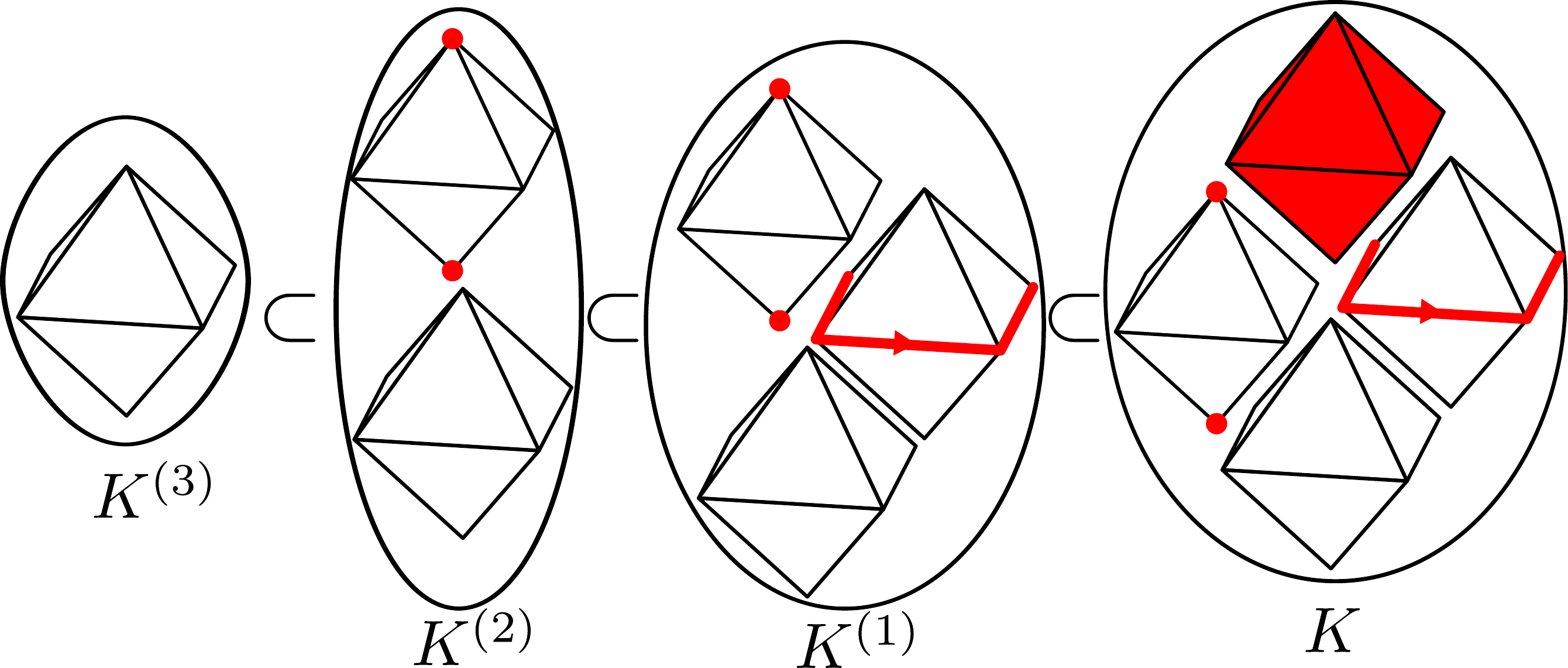}
		\caption{The K-theory classifying group $K$ classifies all bulk
			band structures, regardless their boundary signature. Refined classification
			groups $K^{(n)}$ are defined by excluding topological
			phases with anomalous boundary states of order
			$\le n$. The figure illustrates this procedure for a
			crystal with inversion symmetry $\mathcal{I}$,
showing generators only.
		Anomalous boundary states are indicated in red.
}
		\label{fig:10}
	\end{center}
\end{figure}

The K-theory classification classifies topological band structures without considering boundary signatures. In general, the classifying group $K$ for a given combination of non-spatial and crystalline symmetries contains first-order topological phases, which do not rely on the presence of the crystalline symmetries for their protection, higher-order topological phases, as well as atomic-limit phases that do not have protected boundary states, but may or may not have boundary charges. 
To obtain a boundary-resolved classification, Ref.~\onlinecite{trifunovic2019} proposes to consider a subgroup sequence
\begin{equation}
  K^{(d)} \subset K^{(d-1)} \subset \ldots \subset K^{(1)} \subset K,
  \label{eq:subgroup}
\end{equation}
where $K^{(n)}$ contains those elements of $K$ that do {\em not} have intrinsic boundary signatures of order $n$ or lower, see Fig.~\ref{fig:10} for a schematic illustration. One verifies that $K^{(n)}$ is indeed a subgroup of $K$, since the ``addition'' of band structures ({\em i.e.}, taking the direct sum) can not lower the order of the boundary signatures. In the language used above, this conclusion follows from the observation that under taking direct sums the number of boundary mass terms does not decrease. The quotient $K^{(n+1)}/K^{(n)}$ classifies topological crystalline band structures with exactly $n$ boundary mass terms on. Many of the examples discussed above are generators of the topological classes in these quotient groups.

To illustrate the use of the boundary-resolved
classification~(\ref{eq:subgroup}), we give the subgroup sequences for
one-dimensional inversion-symmetric odd-parity superconductors, of which the
``Kitaev chain'' (\ref{eq:Kitaev}) is an example,
\begin{equation}
  2 \ZZ \subset \ZZ
\end{equation}
and for two-dimensional superconductors with fourfold rotation symmetry ${\cal
R}_4$, 
\begin{equation}
  \ZZ \times 2\ZZ \subset \ZZ^2 \subset \ZZ^3,
\end{equation}
of which the BBH model~(\ref{eq:HBBH}) is an example. These equations
summarize the discussions of the boundary resolved
classifications of the last paragraphs of Secs.~\ref{sec:3.1}
and~\ref{sec:3.2}, respectively.

\section{Bulk-boundary correspondence for topological crystalline phases} \label{sec:3}
The bulk-boundary correspondence relates the topological classification of
anomalous boundary states to the boundary-resolved
classification~(\ref{eq:subgroup}) of the bulk band structure. It states that
(i)
\begin{align}
	{\cal K}_\text{a}^{(n)} =\, K^{(n-1)}/K^{(n)},
	\label{eq:bb}
\end{align}
where ${\cal K}_\text{a}^{(n)}$ is the classification group of anomalous 
$n$th-order boundary states and $K^{(n)}$ is the K-theory classification 
group of topological band structures without intrinsic boundary signature of
order $\le n$, and that (ii) the topological crystalline band structures
without anomalous boundary signature, which are classified by $K^{(d)}$, can be
continuously deformed to atomic-limit phases. Reference~\onlinecite{trifunovic2019}
derives these relations using algebraic methods for order-two crystalline
symmetries, crystalline symmetries that square to one. Such a general
derivation is possible because of the existence of a complete K-theory
classification in this case~\cite{shiozaki2014}. We will discuss a heuristic
derivation of the bulk-boundary correspondence for general point group $G$ at
the end of this Section.

The formal definition of the classifying group ${\cal K}_{\rm a}$ of anomalous
higher-oder boundary states and a method to compute it are presented in
Sec.~\ref{sec:4.1}. The right-hand side of the bulk-boundary
correspondence~(\ref{eq:bb}) contains the subgroups $K^{(n)}$ of the
classifying group of topological crystalline band structures $K$ that classify
the higher-order band structures. Such a refined bulk classification is
obtained by an extension of the method introduced by Cornfeld and
Chapman~\cite{cornfeld2019}, as explained in Sec.~\ref{sec:4.2}. A
constructive proof of the bulk-boundary correspondence~(\ref{eq:bb}) consists
of independent calculations of the left- and the right-hand side of
Eq.~(\ref{eq:bb}) for a given symmetry group. Here, one needs not only
demonstrate a one-to-one correspondence between the two groups in
Eq.~(\ref{eq:bb}), but also that the generators of the groups
$K^{(n-1)}/K^{(n)}$ have the corresponding anomalous boundary states once
terminated.  Since the presence or absence of anomalous boundary states is a
topological property, this verification can be performed for a convenient
choice of the bulk band structure and the termination, such as the low-energy
Dirac-like Hamiltonians with smooth terminations, for which we can use the
domain-wall picture of Sec.~\ref{sec:DW}.

\subsection{Anomalous boundary states} \label{sec:4.1}
\begin{figure}[t]%
	\begin{center}
		\includegraphics*[width=0.95\columnwidth]{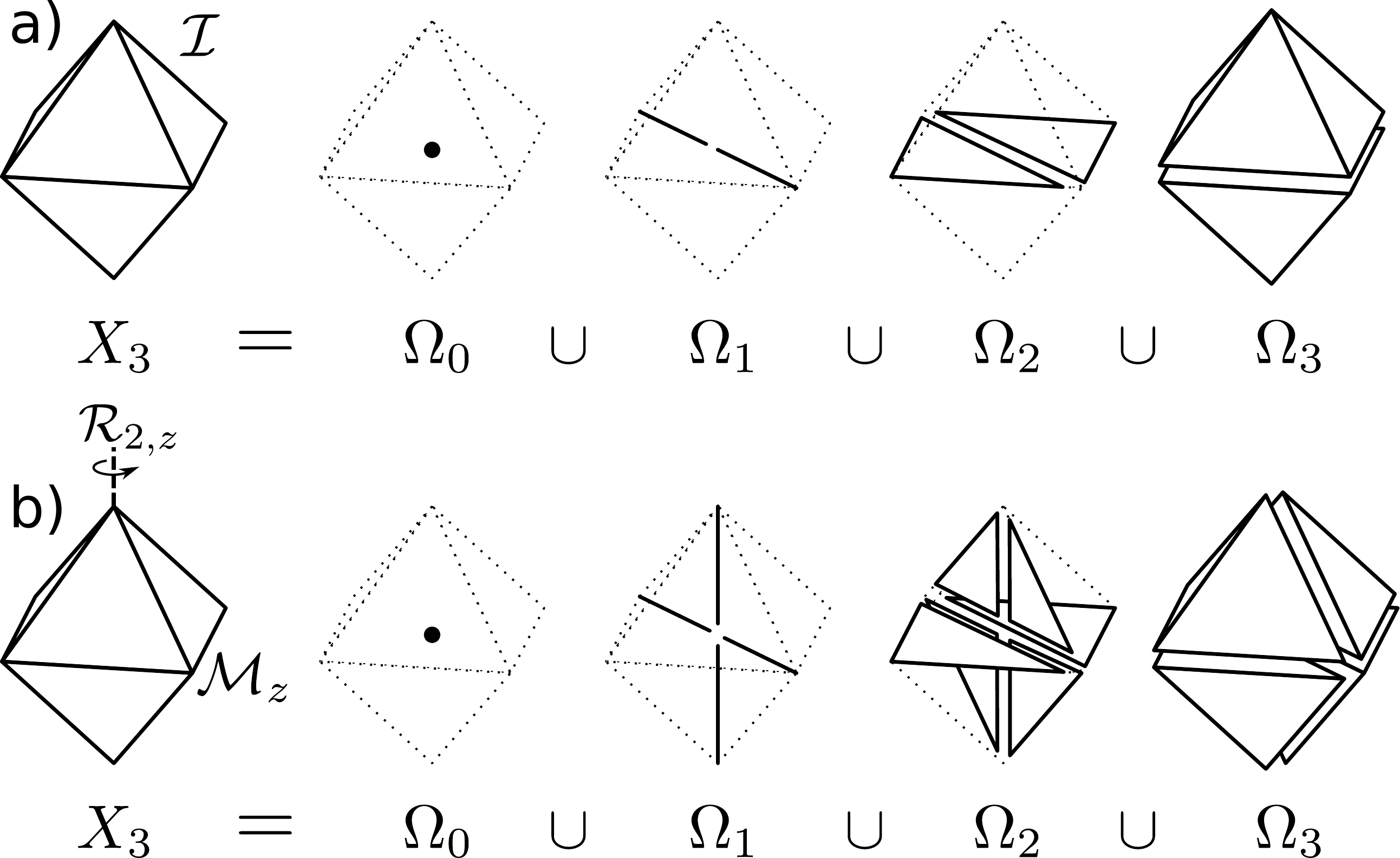}
		\caption{$G$-symmetric cellular decompositions of the octahedron for the crystalline symmetry group $G$ generated by inversion ${\cal I}$ (a) and for $G = C_{2h}$, generated by mirror ${\cal M}_z$ and twofold rotation symmetry ${\cal R}_{2,z}$. }
		\label{fig:6}
	\end{center}
\end{figure}

The classification group ${\cal K}_\text{a}^{(n)}$ classifies anomalous
$n$th-order boundary states for a crystal shape that is compatible with the
crystalline symmetry group $G$. We use the convention that the sum of an
$n$th-order boundary state and boundary state of order larger than $n$ is
considered a boundary state of order $n$. The precise definition of the group
${\cal K}_\text{a}^{(n)}$ requires the notion of the $G$-symmetric cellular
decomposition of a crystal \cite{shiozaki2018b}: Denoting the interior of the
crystal by $X$, one writes 
\begin{equation}
  X = \Omega_0\cup\Omega_1\cup\dots\cup\Omega_d,
	\label{eq:cell_decomp}
\end{equation}
where $d$ is the spatial dimension and $\Omega_k$ is the a set of disjoint
``$k$-cells'' $c_k$ --- a $k$-cell is a $k$-dimensional subset of $X$ that is
homotopic to the interior or a $k$-dimensional sphere --- which have the
property that each element $g \in G$ either leaves each point in $c_k$
invariant or bijectively maps the $k$-cell $c_k$ to a different $k$-dell
$c_{k'}$. Further, for a pair of $k$-cells $c_k$ and $c_{k'}$ in $\Omega_d$
there is one and precisely one $g \in G$ that maps these cells onto each other.
Examples of $G$-symmetric cellular decompositions are shown in Fig.\
\ref{fig:6} for a crystal with inversion symmetry and for a crystal with mirror
and twofold rotation symmetries.

\begin{figure}[t]
	\begin{center}
		\includegraphics*[width=.6\columnwidth]{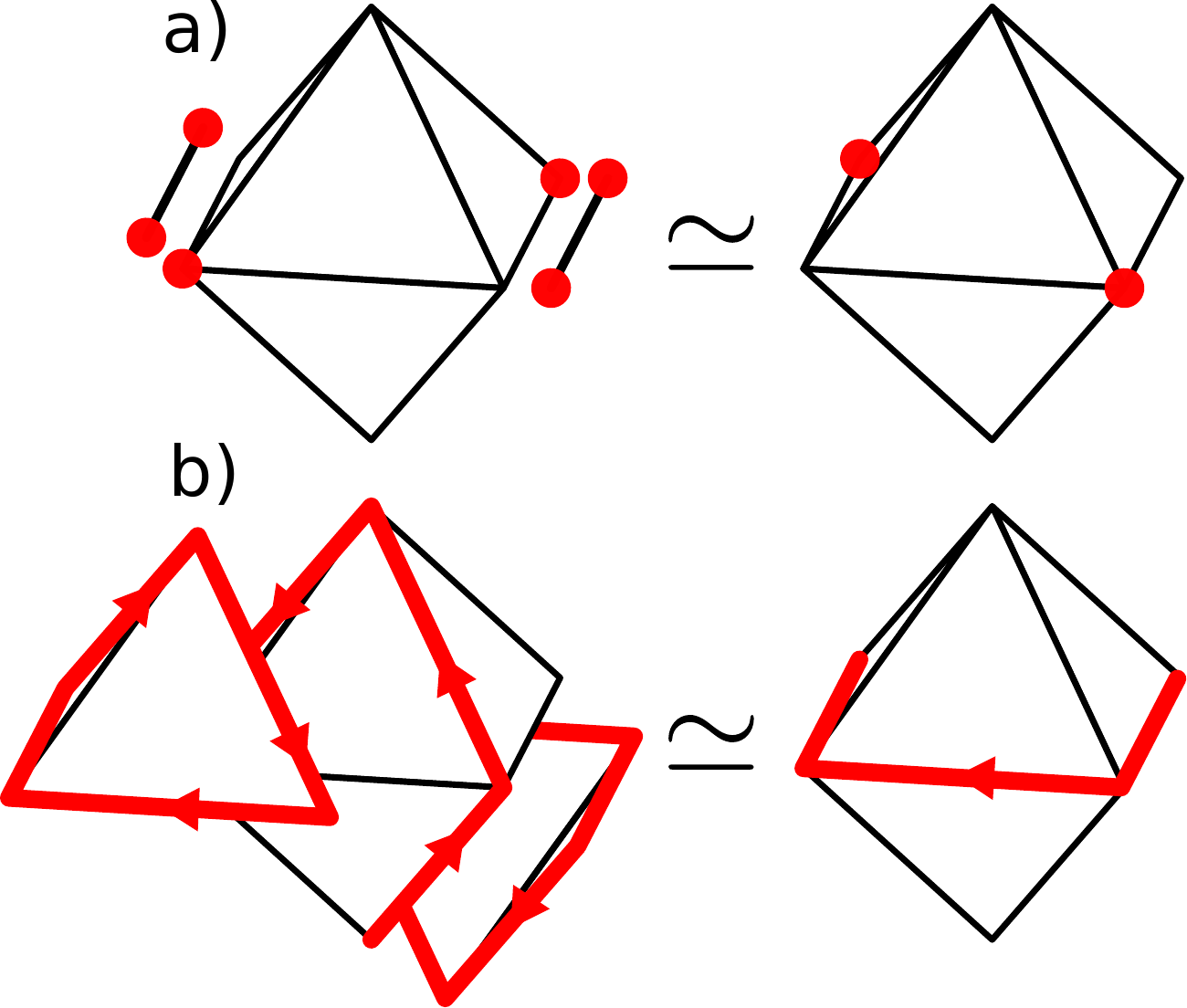}
		\caption{By attaching a ``decoration'' consisting of a
			first-order topological phase on a boundary of
			codimension $n-1$, an arbitrary configuration of
			boundary state of codimension $n$ can be moved to the
			subset $\partial\Omega_{d+1-n}\cap \partial X$. The
			figure shows two examples for a three-dimensional
			crystal with inversion symmetry: Corner states at
			generic corners can be moved to $\partial\Omega_{1}$ by
			changing the crystal termination along two crystal
			hinges (a) and hinge states at a generic hinge can be
			moved to $\partial \Omega_{2}$ by changing the crystal
			termination at two crystal faces (b). In both cases,
			the hinges or faces at which the termination is changed
			are related by inversion.}
		\label{fig:7}
	\end{center}
\end{figure}

To construct topological boundary states of dimension $k-1$, we first consider
the allowed $k$-dimensional topological phases with support on the $k$-cell
$c_k$. Since the only relevant symmetries acting within $c_k$ are local ---
recall that each element $g \in G$ either leaves each point in $c_k$ invariant
or it does not act inside $c_k$ --- any topological phase placed on $c_k$
satisfies the standard bulk-boundary correspondence. This establishes a
one-to-one correspondence between topological phases with support on $c_k$ and
boundary states at its boundary $\partial c_k$.  By placing $k$-dimensional
topological phases on $\Omega_{k}$ in a $G$-compatible manner, we can generate
protected $(k-1)$-dimensional states on the crystal boundary $\partial X$,
provided any topological boundary states that arise in the interior of the
crystal mutually gap out. This procedure gives a construction of all
topological boundary states on $\partial \Omega_{k} \cap \partial X$, both
extrinsic and intrinsic. Since its ``building blocks'', topological phases
defined on the $k$-cells, have a classification with a well-defined group
structure, the result of this procedure has a group structure, too. Setting
$k=d+1-n$, we refer to it as ${\cal K}^{(n)}$, the classifying group of all
$n$-th order topological boundary states on $\partial \Omega_{d+1-n} \cap
\partial X$. Note that, in principle, codimension-$n$ boundary states may also
appear outside $\partial \Omega_{d+1-n}$, but such states can be moved to
$\partial \Omega_{d+1-n}$ by a suitable change of crystal termination along
$d+1-n$-dimensional crystal faces, see Fig.~\ref{fig:7}. 

\begin{figure}[t]%
	\begin{center}
		\includegraphics*[width=.8\columnwidth]{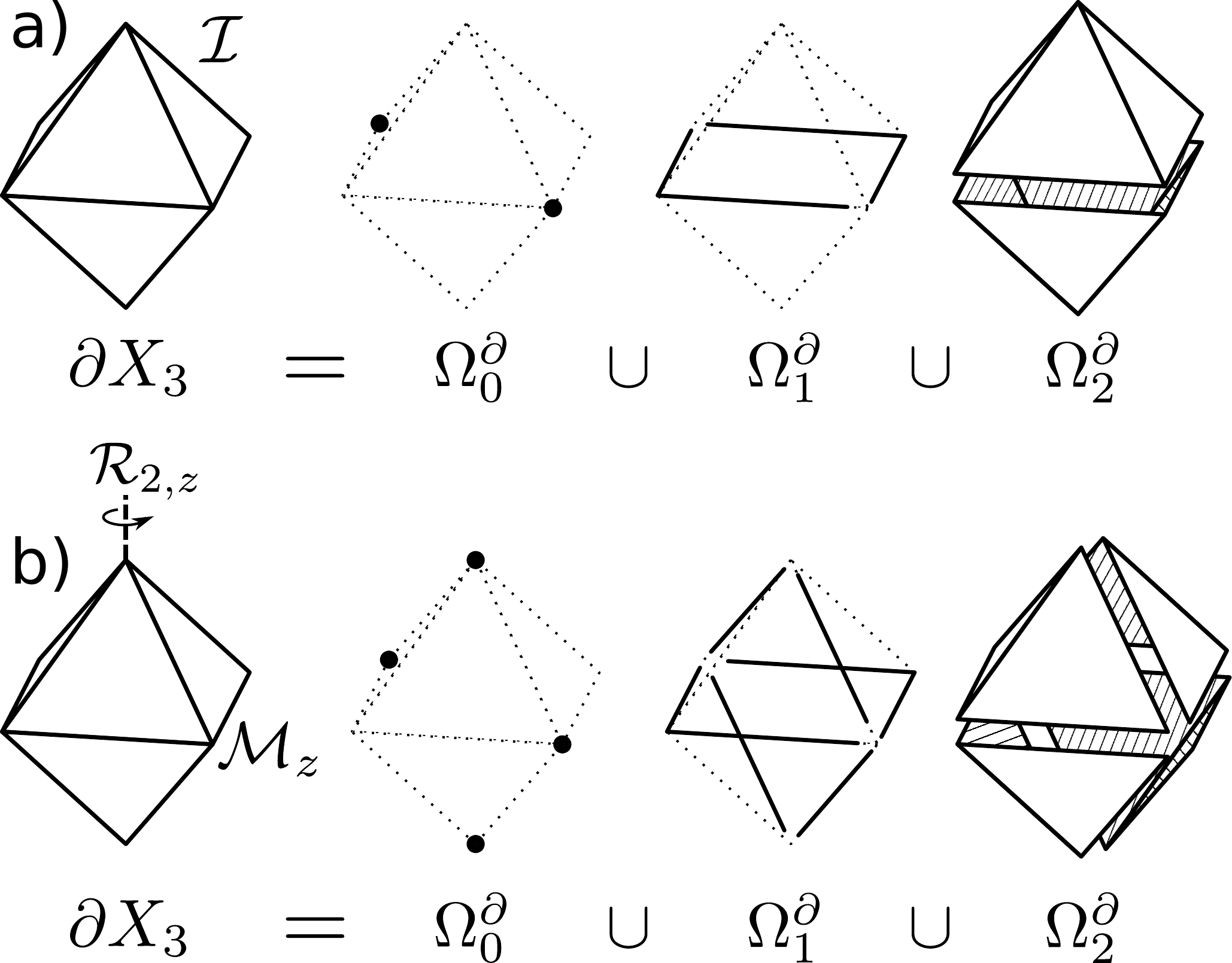}
		\caption{$G$-symmetric cellular decompositions of the boundary
			of an octahedron for $G$ generated by inversion ${\cal I}$ (a) and for $G=C_{2h}$
		generated by mirror- and twofold rotation symmetry.
}
		\label{fig:8}
	\end{center}
\end{figure}
Let us now apply the above general considerations for the example of the point
groups $G = C_i$ (inversion), for which the cellular decomposition is shown in
Fig.~\ref{fig:6}a. (A second example, corresponding to $G = C_{2h}$, see
Fig.~\ref{fig:6}b, will be discussed in Sec.~\ref{sec:5.1}.) The first-order
boundary states are generated and classified by placing a three-dimensional
ten-fold-way phase onto one of the two $3$-cells in $\Omega_3$ (provided
nontrivial phases for the ten-fold-way class of interest). An inverted copy of
this phase on the other $3$-cell, which ensures that the crystal as a whole is
inversion-symmetric. The resulting phase has inversion-symmetric first-order
boundary states if the pair of surface states in the interior of the crystal
can be mutually gapped-out. Similarly, to generate all second-order boundary
states, we place a two-dimensional tenfold-way phase and an inverted copy on
the two $2$-cells in $\Omega_2$. This way an order-two boundary state is
obtained, provided the two one-dimensional states along the interior boundary
between the two $2$-cells gap out. Third-order boundary states are constructed
analogously by placing tenfold-way phases onto the two $1$-cells in $\Omega_1$. 

This construction gives the classifying group ${\cal K}^{(n)}$ of {\em all}
$n$-th order topological boundary states, both intrinsic and extrinsic.
To find the classifying group ${\cal K}_{\rm a}^{(n)}$ of anomalous ({\em
i.e.}, intrinsic) $n$-th order boundary states, one has to divide out the
subgroup ${\cal D}^{(n)} \subset {\cal K}^{(n)}$ of extrinsic boundary states,
{\em i.e.}, of topological states on $\partial \Omega_{d+1-n} \cap \partial X$
that appear as the boundary states of topological phases with support entirely
within the crystal boundary $\partial X$. To classify such states that can be
obtained by ``decoration'' of the crystal boundary $\partial X$, we make use of
the induced $G$-symmetric cellular decomposition of the boundary $\partial X$,
\begin{align}
	\partial X =&\, \Omega^\partial_0\cup\Omega_1^\partial\cup\dots\cup\Omega_{d-1}^\partial,
  \label{eq:Xomega}
\end{align}
where $\Omega_k^\partial=\Omega_{k+1}\cap\partial X$. Figure~\ref{fig:8} shows
the induced $G$-symmetric cellular decomposition of the crystal boundaries for
the examples discussed above. Proceeding as before, all boundary states
classified by  ${\cal D}^{(n)}$ can be obtained by ``pasting'' non-trivial
topological phases (with the appropriate local symmetries, if applicable) onto
$k$-cells $\Omega_k^\partial$ with $k\ge d+1-n$, with the requirement that all
the states of dimension $>d-n$ can be gapped out. The classifying group ${\cal
K}_{\rm a}^{(n)}$ of anomalous $n$th-order boundary states is then
\begin{equation}
  {\cal K}_\text{a}^{(n)}={\cal K}^{(n)}/{\cal D}^{(n)}.
	\label{eq:Ka}
\end{equation}

The importance of considering decorations by phases of dimension larger than
$d+1-n$ is that one should also consider decorations with higher-order
topological phases with support on the crystal boundary.  Reference~\onlinecite{trifunovic2019} discusses an example for which this is relevant: A
three-dimensional inversion-symmetric time-reversal invariant superconductor.
For this example, the classifying group ${\cal K}^{(3)} = \ZZ_2$ is generated
by a boundary state consisting of two Majorana Kramers pairs positioned on the
boundary $0$-cell $\Omega_0^{\partial}$, see Fig.~\ref{fig:8}a. Such a
configuration of Majorana-Kramers pairs can also be obtained from a stand-alone
two-dimensional ${\cal T}$-symmetric superconductor with support on the crystal
surface: In the $G$-symmetric cellular decomposition of the boundary, the two
$2$-cells in $\Omega_2^\partial$ host a two-dimensional ${\cal T}$-symmetric
superconductor with helical Majorana modes. The two helical Majorana modes gap
out at the interface between the two $2$-cells, leaving behind Kramers-Majorana
zero modes at two inversion-related corners.

To make this construction more specific, we will apply it to the example
of an odd-parity topological superconductor with crystalline symmetry
group $C_{2h}$ in Sec.~\ref{sec:5}.

\subsection{Refined bulk classification} \label{sec:4.2}
Close to a phase transition between two different topological phases, the
low-energy description of the band structure can be chosen to take the Dirac
form~(\ref{eq:HdG}).  Since we are interested in classifying strong phases that
are protected by the point-group symmetry $G$, we may assume that the gap
closing appears in the center of the Brillouin zone. For this reason, the
(strong) topological classification of band structures is the same as the
classification of Dirac Hamiltonians.

The key simplifying observation is that close to the phase transition, the
low-wavelength description of the band structure is very ``symmetric'' and
allows the classification problem with point-group symmetries to be mapped to a
classification problem with local ({\em i.e.}, onsite) symmetries or
antisymmetries only~\cite{shiozaki2014,thorngren2018,cornfeld2019}. With an
onsite symmetry group $G$ the Dirac Hamiltonian can be block-diagonalized,
where each ``block'' is labeled by an irreducible representations of $G$. The
individual blocks are no longer constrained by point-group symmetries but only
by non-spatial symmetries that have same mathematical form as the fundamental
symmetries ${\cal P}$, ${\cal T}$ and/or ${\cal C}$. Therefore, each block
corresponds to one of the tenfold-way classes, the classification of which is
well known~\cite{kitaev2009,schnyder2009}. The realization of a mapping between
point-group symmetries and onsite symmetries was first provided by Shiozaki
and Sato for order-two symmetries~\cite{shiozaki2014}, and later extended by
Cornfeld and Chapman to all point-group symmetries~\cite{cornfeld2019}. Below
we refer to this mapping as the ``Cornfeld-Chapman isomorphism''.

Once the topological classification $K$ of gapped Dirac Hamiltonians is known,
the next task is to refine this classification according to the boundary
signatures of the different phases~(\ref{eq:subgroup}). This requires finding
mass terms that are incompatible with the crystalline symmetry group $G$ and
evaluating the order of the boundary state associated with their transformation
behavior under $G$. A complication is that there may be multiple choices for
the mass terms with different numbers of mass terms and/or different order of
the associated boundary signatures. If this complication occurs, the difference
of boundary signatures that correspond to different choices of the mass
terms is {\em extrinsic}, {\em i.e.}, it is associated with a topological
trivial bulk. 

To find a proper correspondence between a Dirac Hamiltonian and its boundary
signature, the configuration of mass terms with the maximal order of the
boundary states has to be found. For simple examples this is most easily
accomplished by inspection of the Dirac Hamiltonians corresponding to the
generators of $K$, as was done in the examples discussed in Sec.~\ref{sec:2}.
We now discuss a systematic procedure that gives the same result.

The calculation of the subgroup $K^{(n)}$ of topological phases with boundary
signature of order not lower than $n$ proceeds in two steps. First one selects
all $n$-dimensional real (but not necessarily irreducible) representations
$O_n$ of the point group $G$ that correspond to $(n+1)$th-order boundary
signatures if $n$ mass terms $M_1$, \ldots, $M_n$ were to transform under $G$
with the representation $O_n$. Second, the classification group $K_{O_n}$ for
Dirac Hamiltonians with mass terms that transform under $G$ in this manner is
obtained. Since Dirac Hamiltonians in $K_{O_n}$ satisfy the full crystalline
symmetry group $G$ if the mass terms $M_1$, \ldots, $M_n$ are omitted, there is
a natural inclusion $K_{O_n} \hookrightarrow K$. The subgroup $K^{(n)} \subset
K$ is generated by the group $K^{(n+1)}$ and by the images
$K_{O_n}\hookrightarrow K$, from all the representations selected in the first
step.  

The above procedure is simplified by the following two observations: First,
the calculation of $K^{(d)}$, which is needed as a starting point, is achieved
using the observation made by Shiozaki~\cite{shiozaki2019}, that it is
sufficient to consider only a single representation of the point group, the
``vector representation'' $O_d^\text{vec}$, in which the mass terms $M_1$,
\ldots, $M_d$ transform in the same way as the position vector. Shiozaki proved
this statement by showing that $K_{O_d^\text{vec}}$ is isomorphic to the
classification of zero-dimensional phases placed onto $\Omega_0$, taking into
account the symmetry restrictions imposed by the full crystalline symmetry
group, which acts onsite on $\Omega_0$. Since all atomic-limit phases, in
presence of wallpaper or space group constraints, are obtained by placing
zero-dimensional phases on different Wyckoff
positions~\cite{trifunovic2019,po2017,geier2019}, which for the case of
point-group symmetries reduces to a single position $c_0\in\Omega_0$, one
concludes that the image of $K_{O_d^\text{vec}}\hookrightarrow K$ is precisely
$K^{(d)}$. For example, a two-dimensional system with mirror symmetry does not
have topologically non-trivial atomic limits since $\Omega_0$ is empty, while
for three-dimensional system with inversion symmetry the non-trivial atomic
limits are obtained by placing a zero-dimensional inversion-symmetric
Hamiltonian on the inversion-symmetric point in $\Omega_0$, see
Fig.~\ref{fig:6}a. The second observation concerns the calculation of
$K^{(1)}$, which classifies Dirac Hamiltonians that are trivialized once the
constraints posed by the point group symmetries are lifted. This group is most
easily computed as the kernel of  the inclusion $K\hookrightarrow K_\text{TF}$,
where $K_\text{TF}$ is the tenfold-way classifying group without point-group
constraints. This way, the bulk subgroup sequence~(\ref{eq:subgroup}) is
readily obtained for $d=1$, $2$. In three spatial dimensions, the only
additional task that needs to be accomplished is the computation of the
subgroup $K^{(2)}$, which requires considering all two-dimensional real
representations $O_2$ corresponding to third-order boundary signatures. 

\subsection{Heuristic proof of bulk-boundary correspondence} 
The statement that the bulk classifying group $K^{(d)}$ of topological
crystalline band structures without anomalous boundary signatures precisely
describes the atomic-limit phases follows from the observation that the absence
of anomalous boundary signatures implies that a crystal can be smoothly ``cut''
into smaller sub-units without generating gapless modes at interfaces.
Performing this cutting procedure in a periodic manner gives the desired
connection to an atomic-limit phase.

For the proof of the relation~(\ref{eq:bb}) the only nontrivial part is the
surjectivity of the map $K^{(n-1)}/K^{(n)} \to {\cal K}^{(n)}_{\rm a}$.  In
this regard, we note that the construction of the classifying group ${\cal
K}^{(n)}$ in Sec.~\ref{sec:4.1} entails that all (anomalous) states on the
boundary of a $G$-symmetric crystal $X$ can be obtained by
``embedding''~\cite{tuegel2019} $(d+1-n)$-dimensional topological phases onto
$(d+1-n)$-cells from the element $\Omega_{d+1-n}$ of its $G$-symmetric cellular
decomposition.  It remains to be demonstrated that such ``embedded'' phases can
be made translationally invariant. For order-two symmetries, this was proven
with an algebraic method that constructs topological crystalline band
structures with a Dirac low-energy description for a given element from ${\cal
K}_\text{a}^{(n)}$ \cite{trifunovic2019}. Alternatively, for order-two
symmetries one can invoke a layer-stacking
construction~\cite{isobe2015,fulga2016}. For an arbitrary symmetry group $G$,
Refs.~\onlinecite{huang2017,song2019b} propose the ``topological crystal
construction''~\cite{huang2017}, in which $X$ is viewed as a $G$-symmetric
unit-cell of a periodic lattice, which is repeated periodically in space. In
comparison to the construction of Sec.~\ref{sec:4.2} this construction imposes
one additional condition: Boundary states at the ``seam'' between neighboring
unit-cells have to be gapped out. To the best of our knowledge, there is no
proof that this condition is always met for an arbitrary symmetry group $G$,
although we do not know of any counter examples.

Alternatively, the bulk-boundary correspondence~(\ref{eq:bb}) can
be proven by independently calculating the bulk subgroup sequence and boundary
classification using the methods reviewed in Secs.~\ref{sec:4.1} and \ref{sec:4.2}. In the
following Section we illustrate this procedure on one example.

\section{Example: Odd-parity superconductor.}
\label{sec:5}
To make the construction of Secs.~\ref{sec:4.1} and \ref{sec:4.2} more
explicit, we apply it to the example of an odd-parity topological
superconductor with crystalline symmetry group $C_{2h}$ (the point group
generated by mutually commuting twofold rotation symmetry ${\cal R}_{2,z}$ and
mirror symmetry ${\cal M}_z$). Using the convention ${\cal R}_{2,z}^2 = {\cal
M}_z^2 = 1$, ${\cal M}_{z}$ anticommutes with particle-hole conjugation ${\cal
P}$, whereas and ${\cal R}_{2,z}$ commutes with ${\cal P}$. An eight-band Dirac
Hamiltonian for this symmetry class was discussed and analyzed in
Sec.~\ref{sec:boundary_dirac}.

Below we carry out the constructive proof of the bulk-boundary
correspondence~(\ref{eq:bb}). In Sec.~\ref{sec:5.1} we first obtain the
left-hand side of Eq.~(\ref{eq:bb}), the classification of anomalous
higher-order boundary states. In Sec.~\ref{sec:5.2} we subsequently compute
the classification group $K$ of topological crystalline bulk band structures
and its subgroups $K^{(n)}$, which classify higher-order band structures.  The
bulk classification provides generators of Dirac-like form, for which we find
the anomalous boundary signatures using domain-wall picture. Together, these
three steps give a constructive proof of the bulk-boundary
correspondence~(\ref{eq:bb}) for this symmetry class. An additional example in
two dimensions can be found in the appendix.

\subsection{Boundary classification}
\label{sec:5.1}

We first show that the boundary classification for this example is
\begin{equation}
  {\cal K}^{(1)}_{\rm a} = 0,\ \
  {\cal K}^{(2)}_{\rm a} = 0,\ \
  {\cal K}^{(3)}_{\rm a} = \ZZ_2.
  \label{eq:boundary_example}
\end{equation}
Calculation of the boundary classification groups requires the $G$-symmetric cellular decompositions of the crystal and the crystal boundary, which are shown in Figs.\
\ref{fig:6}b and \ref{fig:8}b, respectively. 
The result for ${\cal K}^{(1)}$ follows immediately, as there are no first-order boundary states --- there are no non-trivial three-dimensional superconductors that can be placed onto $3$-cells in $\Omega_3$. 

\begin{figure}[t]%
	\begin{center}
		\includegraphics*[width=0.9\columnwidth]{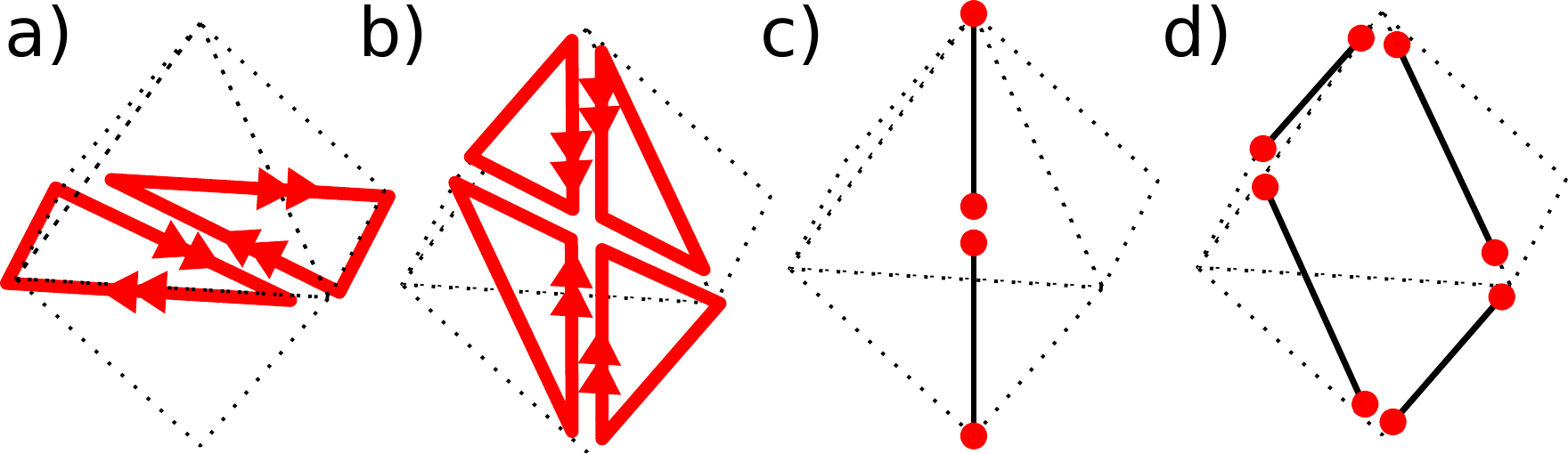}
		\caption{Construction of second-order boundary states for a
			crystal with ${\cal M}_z$ and ${\cal R}_{2,z}$
			symmetries by placing two-dimensional topological
			superconductors on the two $2$-cells in the mirror
			plane (a). Placing two-dimensional superconductors on
			the four $2$-cells perpendicular to the mirror plane is
			not allowed, because the co-propagating chiral Majorana
			modes at the interior boundaries between cells can not
			be gapped out (b). Third-order boundary states with
			well-defined ${\cal R}_{2,z}$ parity can be obtained by
			placing one-dimensional superconductors at the two
			$1$-cells along the rotation axis (c). Extrinsic
			third-order boundary states, consisting of pairs of
			corner states with different ${\cal R}_{2,z}$ parity,
		arise from one-dimensional topological phases at the four
	boundary $1$-cells outside the mirror plane (d).}

		\label{fig:example}
	\end{center}
\end{figure}

Second-order boundary states may be obtained by placing two-dimensional
topological superconductors with chiral Majorana edge modes on the two
$2$-cells in the mirror plane or on the four $2$-cells perpendicular to the
mirror plane, see Figs.~\ref{fig:example}a and b. In the former case, the local
${\cal M}_z$ symmetry imposes that the number of chiral Majorana modes of each
of the two topological phases is even, since ${\cal M}_z$ anticommutes with
${\cal P}$. At the interior ``seam'' between the two $2$-cells in the mirror
plane the Majorana modes are counterpropagating and can be gapped out. Hence,
placing topological phases at the $2$-cells in the mirror plane results in a
topological boundary state consisting of two co-propagating chiral Majorana
modes in the mirror plane. The classifying group of such boundary states is
$2\ZZ$, the same as the classification group of two-dimensional topological
superconductors with an onsite symmetry anticommuting with ${\cal P}$
\cite{trifunovic2019}. Placing two-dimensional topological superconductors with
chiral Majorana edge modes on the four $2$-cells perpendicular to the mirror
plane does not result in a valid topological boundary state, since the chiral
Majorana modes are co-propagating at the rotation axis and cannot be gapped out
there, see Fig.~\ref{fig:example}b. We thus conclude that for this example
${\cal K}^{(2)} = 2\ZZ$.

To construct third-order boundary states, we place one-dimensional topological
superconducting phases on the two $1$-cells along the rotation axis, see Fig.\
\ref{fig:example}c. Because of the presence of the local ${\cal R}_{2,z}$
symmetry commuting with ${\cal P}$, such one-dimensional topological
superconducting phases have a $\ZZ_2^2$ classification, since the Majorana end
states have well-defined ${\cal R}_{2,z}$-parity. Since ${\cal M}_z$
commutes with ${\cal R}_{2,z}$, the two mirror-related Majorana states at the
crystal center have the same ${\cal R}_{2,z}$-parity and can mutually gap out.
There are no one-dimensional topological superconductors with an onsite
symmetry anticommuting with ${\cal P}$ \cite{trifunovic2019}, 
so placing topological phases onto the
two $1$-cells in the mirror plane is not possible. We thus conclude that ${\cal
K}^{(3)} = \ZZ_2^2$, corresponding to Majorana corner states on the rotation
axis with even or odd ${\cal R}_{2,z}$-parity.

To find the decoration group ${\cal D}^{(2)}$, we consider two-dimensional
topological superconductors placed on the eight faces of $\partial \Omega_2$, see
Fig.~\ref{fig:5}c. The counter-propagating Majorana modes at the ``seams''
between boundary $2$-cells outside the mirror plane are gapped out,\footnote{There is a subtlety here: The surface decoration generating ${\cal
D}^{(2)}$ also contains a corner state at each of the two ${\cal
R}_{2,z}$-symmetric corners, see Fig.~\ref{fig:5}c. This can be verified by
explicit calculation or, alternatively, from the discussion of this example in
Sec.~\ref{sec:boundary_dirac}, where it is shown that a boundary signature with
two co-propagating Majorana modes in the mirror plane but without a corner
state in the two ${\cal R}_{2,z}$-symmetric corners is anomalous. As a
consequence, using the decoration from ${\cal D}^{(2)}$, the hinge states can
be gapped out, although by doing so the corner states appear.} while the pairs
of co-propagating Majorana modes in the mirror plane remain. We conclude that
${\cal D}^{(2)} = 2\ZZ$ and, hence, ${\cal K}_{\rm a}^{(2)} = 0$. 

To find ${\cal D}^{(3)}$ we add one-dimensional topological superconductors at
the four boundary $1$-cells outside the mirror plane, see Fig.\
\ref{fig:example}d. The pairs of Majorana end states in the mirror plane can
mutually gap out because ${\cal M}_z$ anticommutes with ${\cal P}$. What
remains are pairs of Majorana corner modes at the two ${\cal
R}_{2,z}$-symmetric corners, one of each ${\cal R}_{2,z}$ parity. No
one-dimensional topological superconductors can be pasted onto the four
boundary $1$-cells in the mirror plane, because the existence of a local
symmetry anticommuting with ${\cal P}$ rules out topological phases there. We
conclude that ${\cal D}^{(3)} = \ZZ_2$ and, hence, ${\cal K}_{\rm a}^{(3)} =
\ZZ_2$.

\subsection{Bulk classification}
\label{sec:5.2}

We now use the general method outlined in Sec.~\ref{sec:4.2} to show that from the bulk perspective the odd-parity superconductor with point group $G = C_{2h}$ is classified by the subgroup sequence
\begin{equation}
  2 \ZZ \subset \ZZ \subset \ZZ \subset \ZZ.
\end{equation}
This subgroup sequence is consistent with the boundary classification (\ref{eq:boundary_example}).

A Dirac Hamiltonian for this class was discussed in Sec.\
\ref{sec:boundary_dirac}. There, the relevant mass terms could easily be found
by inspection. To illustrate the systematic formalism of Sec.~\ref{sec:4.2},
we here rederive the same results using the Cornfeld-Chapman isomorphism.

We first calculate the full classification group $K$ of three-dimensional gapped Dirac Hamiltonians
\begin{align}
	H =&\, m\Gamma_0+k_1\Gamma_1+k_2\Gamma_2+k_3\Gamma_3,
	\label{eq:hdirac}
\end{align}
with particle-hole constraint $U_{\cal P}$ with ${\cal P}^2=1$,
twofold rotation symmetry $U_{\cal R}$ and mirror symmetry $U_{\cal M}$. At
this point no explicit representation of the gamma matrices and of the symmetry
representations $U_{\cal P}$, $U_{\cal R}$, and $U_{\cal M}$ needs to be
specified.  The Dirac matrices $\Gamma_i$ provide a representation
$U^\Gamma_{\cal R}=ie^{\Gamma_1\Gamma_2\pi/2}$ of twofold rotation symmetry and
a representation $U^\Gamma_{\cal CM}=\Gamma_3$ of a mirror
\textit{antisymmetry}. The superscript $\Gamma$ denotes that these
representations are constructed from the Dirac Hamiltonian and that they in
general satisfy different algebraic relations than the point group symmetries
${\cal R}_{2,z}$ and ${\cal M}_z$. According to the Cornfeld-Chapman
isomorphism the topological classification of Dirac Hamiltonians with the point
group $G$ is the same as the topological classification of Dirac Hamiltonians
with the onsite symmetries
\begin{align}
	U^{\cal O}_{\cal R} =&\, U_{\cal R}^{\Gamma} U_{\cal R},\ \
	U^{\cal C}_{\cal M} = i U_{\cal CM}^{\Gamma} U_{\cal M}.
	\label{eq:Ulocal}
\end{align}
Here the superscripts ${\cal O}$ and ${\cal C}$ indicate the representation
acts as onsite symmetry or antisymmetry ({\em i.e.}, chiral constraint),
respectively. 

The local symmetries ${\cal O}_{\cal R}$ and ${\cal C}_{\cal M}$ obtained
in this manner have different algebraic relations to ${\cal P}$ than the 
original symmetries ${\cal R}$ and ${\cal M}$, although they still mutually commute: ${\cal O}_{\cal R}$
anticommutes with ${\cal P}$, whereas ${\cal C}_{\cal M}$ commutes with
${\cal P}$. By considering a basis with well defined parity $\pm$ under 
${\cal O}_{\cal R}$, the Dirac Hamiltonian is block-diagonalized,
$H = \mbox{diag}\, (h_+, h_-)$. Since ${\cal P}$ anticommutes with ${\cal O}_{\cal R}$, $h_+$ and $h_-$ are related to each other by ${\cal P}$, so that it is sufficient to classify the even-parity 
block $h_+$. Since the only constraint on this block is the antisymmetry 
${\cal C}_{\cal M}$, it belongs to the tenfold way class AIII, 
which has a $K=\ZZ$ classification in three spatial dimensions
\cite{kitaev2009,schnyder2009}. To find the 
corresponding topological invariant, one has to write $h_+$
in the Dirac form similar to Eq.~(\ref{eq:hdirac}), with gamma 
matrices $\gamma_i$, $i=0,1,2,3$, and find the representation 
$u^{\cal C}_{\cal M}$ of the onsite antisymmetry within this block. The topological invariant ${\cal N}$ then reads \cite{shiozaki2019}
\begin{equation}
  {\cal N} = {\frac{1}{4}} \mathrm{tr}\, \gamma_0 \gamma_1 \gamma_2 \gamma_3 u^{\cal C}_{\cal M}.
	\label{eq:topinv}
\end{equation}

For the calculation of the subgroup $K^{(3)}$ classifying the atomic
limit phases, we consider Dirac Hamiltonians with three mass terms $M_{i}$,
$i=1,2,3$, and the three-dimensional vector representation $O^{\text{vec}}_3({\cal R}_{2,z})=-O^{\text{vec}}_3({\cal M}_z)=\mathrm{diag}(-1,-1,1)$ of the
point group. The classification group $K_{O^{\text{vec}}_3}$, classifies
an extended Dirac Hamiltonian with three ``defect coordinates'' $x$, $y$, and $z$,
\begin{align}
	H_{O^{\text{vec}}_3}&=H+x M_1+y M_2+z M_3,
	\label{eq:Hhedge}
\end{align}
where $H$ is the Dirac Hamiltonian given in Eq.~(\ref{eq:hdirac}).
Similarly as before, we construct a representation $\tilde U_{\cal
R}^\Gamma=e^{\Gamma_1\Gamma_2\pi/2}e^{M_1M_2\pi/2}$ of twofold rotation
symmetry and $\tilde U_{\cal M}^\Gamma=i\Gamma_3M_3$ of mirror \textit{symmetry}
using the gamma matrices and mass terms appearing in Eq.~(\ref{eq:Hhedge}).
Applying the Cornfeld-Chapman isomorphism, we map the problem of classifying the
defect Hamiltonian~(\ref{eq:Hhedge}) with the spatially non-local symmetry 
constraints ${\cal R}$ and ${\cal M}$ to that of the classification with the local symmetry 
constraints 
\begin{align}
	\tilde U^{\cal O}_{\cal R} =\, \tilde U^{\Gamma}_{\cal R} U_{\cal R},\ \
        \tilde U^{\cal O}_{\cal M} = \tilde U^{\Gamma}_{\cal M} U_{\cal M}.
	\label{eq:URlocal}
\end{align}
In this case, the onsite symmetry $\tilde {\cal O}_{\cal R}$ commutes with ${\cal
P}$, whereas $\tilde {\cal O}_{\cal M}$ anticommutes with ${\cal P}$. We
use a basis with well-defined parity under $\tilde{\cal O}_{\cal R}$ to write
$H_{O^{\text{vec}}_3}$ in block diagonal form, $H_{O^{\text{vec}}_3} = \mbox{diag}\, (\tilde h_+,\tilde h_-)$. The two blocks
$\tilde h_{\pm}$ of even/odd parity states are classified independently. Each
of these blocks can again be divided into two subblocks, $\tilde h_{\pm} =
\mbox{diag}\, (\tilde h_{\pm,+},\tilde h_{\pm,-})$, defined according to the
parity under $\tilde {\cal O}_{\cal M}$. The subblocks $\tilde h_{\pm,+}$ and
$\tilde h_{\pm,-}$ are not independent, since they are mapped onto each other
by ${\cal P}$. Hence, only two independent blocks remain, say $\tilde h_{\pm,+}$.
Each of these is in tenfold-way class A, because no further symmetry constraints
apply. As shown by Teo and Kane~\cite{teo2010}, the three-dimensional Dirac
Hamiltonian with three defect dimensions (\ref{eq:Hhedge}) has the same
classification as zero-dimensional Hamiltonians in tenfold-way class A.
Accordingly, we have $K_{O^{\text{vec}}_3}=\ZZ^2$.

The image of the inclusion $K_{O^{\text{vec}}_3}\hookrightarrow K$ is obtained
by calculating the topological invariant (\ref{eq:topinv}) for the two
generators of $K_{O^{\text{vec}}_3}$. At this point, it is helpful to introduce
a concrete realization of the Dirac matrices and mass terms of the generator of
$K_{O^{\text{vec}}_3}$ with even or odd $\tilde {\cal O}_{\cal R}$-parity,
\begin{align}
  \tilde \gamma_0 =&\, -\mu_3\tau_3\sigma_2 \nonumber \\ 
  (\tilde \gamma_1,\tilde \gamma_2,\tilde \gamma_3) =&\,
  -(\tau_3\mu_3\rho_3\sigma_2,\tau_3\mu_3\rho_2,\tau_3\mu_2) \nonumber \\ 
  (\tilde m_1,\tilde m_2,\tilde m_3) =&\,
  (\tau_3\mu_3\rho_3\sigma_1,\tau_3\mu_3\rho_1,\tau_3\mu_1),
\end{align}
with constraints $\tilde u_{\cal P}=\tau_1$, $\tilde u^{\cal O}_{\cal M}=\tau_3$
and $\tilde u^{\cal O}_{ {\cal R}}=\pm 1$. To calculate the image in $K$, we
first transform back to the original formulation with spatial symmetries ${\cal
M}$ and ${\cal R}$ using the inverse of Eq.~(\ref{eq:URlocal}). This gives
$U_{\cal R}=\pm\rho_3\sigma_3$ and $U_{\cal M}=\tau_3\mu_3$. Next, we use
Eq.~(\ref{eq:Ulocal}) to transform to the onsite constraints $U_{\cal R}^{\cal
O}=\pm\rho_2\sigma_1$ and $U^{\cal C}_{\cal M}=\mu_1$ and project onto the
even-parity block $h_+$. Calculating the topological invariant
(\ref{eq:topinv}) then gives ${\cal N} = \pm2$. We conclude that the image of
$K_{O^{\text{vec}}_3}\hookrightarrow K$ is $K^{(3)}=2\ZZ$. 

Finally, to calculate $K^{(2)}$ we need to consider all two-dimensional
representations of two mass terms that correspond to third-order boundary
states for all relevant two-dimensional real representations $O_2$ of $C_{2h}$.
We here consider the representation $O_2({\cal M}_z)=-O_2({\cal
R}_{2,z})=\mathrm{diag}(1,1)$. Other representations are possible, too, and can
be treated with the same formalism, but do not affect our conclusions. The
classifying group $K_{O_2}$ classifies defect Hamiltonians of form
\begin{equation}
	H_{O_{2}}=H+x M_1+y M_2,
	\label{eq:Hhedge2}
\end{equation} 
where $H$ is the Dirac Hamiltonian of Eq.~(\ref{eq:hdirac}). The
Cornfeld-Chapman isomorphism maps the point-group symmetries to local symmetry
representations,
\begin{align}
	\bar U_{\cal R}^{\cal O}=\, e^{\Gamma_1\Gamma_2\pi/2}e^{M_1M_2\pi/2}U_{\cal R},\ \
	\bar U_{\cal M}^{\cal C}=i\Gamma_3 U_{\cal M},
	\label{eq:U2Rlocal}
\end{align}
where $\bar {\cal O}_{\cal R}$ and $\bar {\cal C}_{\cal M}$ commute with ${\cal
P}$ and mutually. Block-diagonalizing $H_{O_{2}}$ according to the $\bar {\cal
O}_{\cal R}$-parity gives two independent blocks $\bar h_{\pm}$, which are
effectively in tenfold-way class BDI because of the local constraints $\bar
{\cal C}_{\cal M}$ and ${\cal P}$. The topological classification of
three-dimensional Dirac Hamiltonians with two defect coordinates $x$ and $y$ is
the same as classification of one-dimensional Dirac
Hamiltonians~\cite{teo2010}, which have the classifying group $\ZZ$ for class
BDI. Since the two blocks with even and odd ${\cal O}_{\cal R}$-parity are
independent, we arrive at the classifying group $K_{O_2}=\ZZ^2$.

To find the image of the inclusion $K_{O_2} \hookrightarrow K$, we start from
an explicit realization of the two generators of a three-dimensional Dirac
Hamiltonian in class BDI with defect dimension two, 
\begin{align}
	\bar \gamma_0 =&\, \mu_3\tau_3\sigma_2,\nonumber\\
	(\bar \gamma_1,\bar \gamma_2,\bar \gamma_3) =&\, (\mu_3\tau_3\sigma_1,\mu_3\tau_3\sigma_3,\mu_3\tau_1),\nonumber\\
	(\bar m_1,\bar m_2) =&\, (\mu_3\tau_2,\mu_2),
	\label{eq:H2gen}
\end{align}
with constraints $\bar u_{\cal P}=1$, $\bar u_{\cal C}=\mu_1$, and $\bar
u^{\cal O}_{\cal R} = \pm 1$. As before, we first map back to the formulation
with the spatial symmetries $U_{\cal M}$ and $U_{\cal R}$ using the inverse of
Eq.~(\ref{eq:U2Rlocal}), which gives $U_{\cal R}=\pm\mu_1\tau_2\sigma_2$ and
$U_{\cal M}=\mu_2\tau_1$, and then map to a formulation with the onsite
constraints $U^{\cal O}_{\cal R} = \pm \mu_1 \tau_2$ and $U^{\cal C}_{\cal M} =
\mu_1$ using Eq.~(\ref{eq:Ulocal}). Finally, we transform to a basis with
well-defined $U^{\cal O}_{\cal R}$-parity and find the topological invariants
${\cal N} = \pm 1$.  From this, we conclude that the $K^{(2)} = \ZZ$.

\section{Conclusion}\label{sec:6}

The discovery of higher-order topological phases~\cite{schindler2018} provided
important insights for the classification of topological crystalline 
insulators and
superconductors. In one set of classification
approaches~\cite{khalaf2018,rasmussen2018,shiozaki2018b,huang2017}, one first
classifies anomalous higher-order boundaries and symmetry-obstructed atomic
limits, and then, assuming bulk-boundary correspondence, finds
the bulk classification group $K$. A different classification paradigm uses
algebraic methods to classify Dirac
Hamiltonians~\cite{chiu2013,morimoto2013,shiozaki2014,cornfeld2019}. Both
approaches provide complete classifications and a set of generating models; the
latter approach provides minimal Dirac models, whereas in the former one uses
the ``topological crystalline construction''~\cite{huang2017} which typically
results in models with a non-minimal unit cell. 

The quest for novel topological materials requires not only the knowledge of
topological classification, but also an efficient method to relate the given
band structure to its boundary signatures. One possibility in this regard is an
algorithm that ``deforms'' a given band structure to a form that is close to a
Dirac-like phase transition. The other possibility, which is actively pursued
by many research groups, concerns the construction of easy-to-compute
topological invariants. If furthermore, such topological invariants are
designed to ``detect'' only band structures with anomalous boundary states, one
refers to these as symmetry-based
indicators~\cite{kruthoff2017,po2017,bradlyn2017}. Symmetry-based indicators
were initially constructed for insulators, and recently extended to
superconductors~\cite{ono2018,ono2019b,ono2019c,shiozaki2019b,skurativska2020,schindler2020,geier2019}.
Although symmetry-based indicators in general do not provide a full
classification, their construction and subsequent application resulted in the
discovery of many new topological insulator
materials~\cite{zhang2019,vergniory2019}. It remains to be seen if
the same will be the case for topological superconductors.

\section{Acknowledgements}
We thank Max Geier, Titus Neupert, and Felix von Oppen for discussions. LT thanks
Ken Shiozaki for useful discussions about Cornfeld-Chapman isomorphism. We
acknowledge support by project A03 of the CRC-TR 183 and by the priority
programme SPP 166 of the German Science Foundation DFG (PWB), as well as by the
Ambizione Grant PZ00P2.179962 of the FNS/SNF (LT).
\appendix

\section{Bulk and bounday classification for two-dimensional superconductors with fourfold rotation symmetry}
\label{sec:app}
In this appendix we consider both the boundary and the bulk classification of a two-dimensional superconductor with fourfold rotation symmetry with $(U_{\cal R})^4=-1$. We 
consider the case that the superconducting order parameter is invariant under
the fourfold rotation symmetry, {\em i.e.}, that $U_{\cal R}$ commutes with particle-hole conjugation ${\cal P}$. The BBH model discussed in the main text is one
example of a topologically non-trivial second-order phase in this class.

\subsection{Boundary classification}

\begin{figure}[t]%
	\begin{center}
		\includegraphics*[width=0.9\columnwidth]{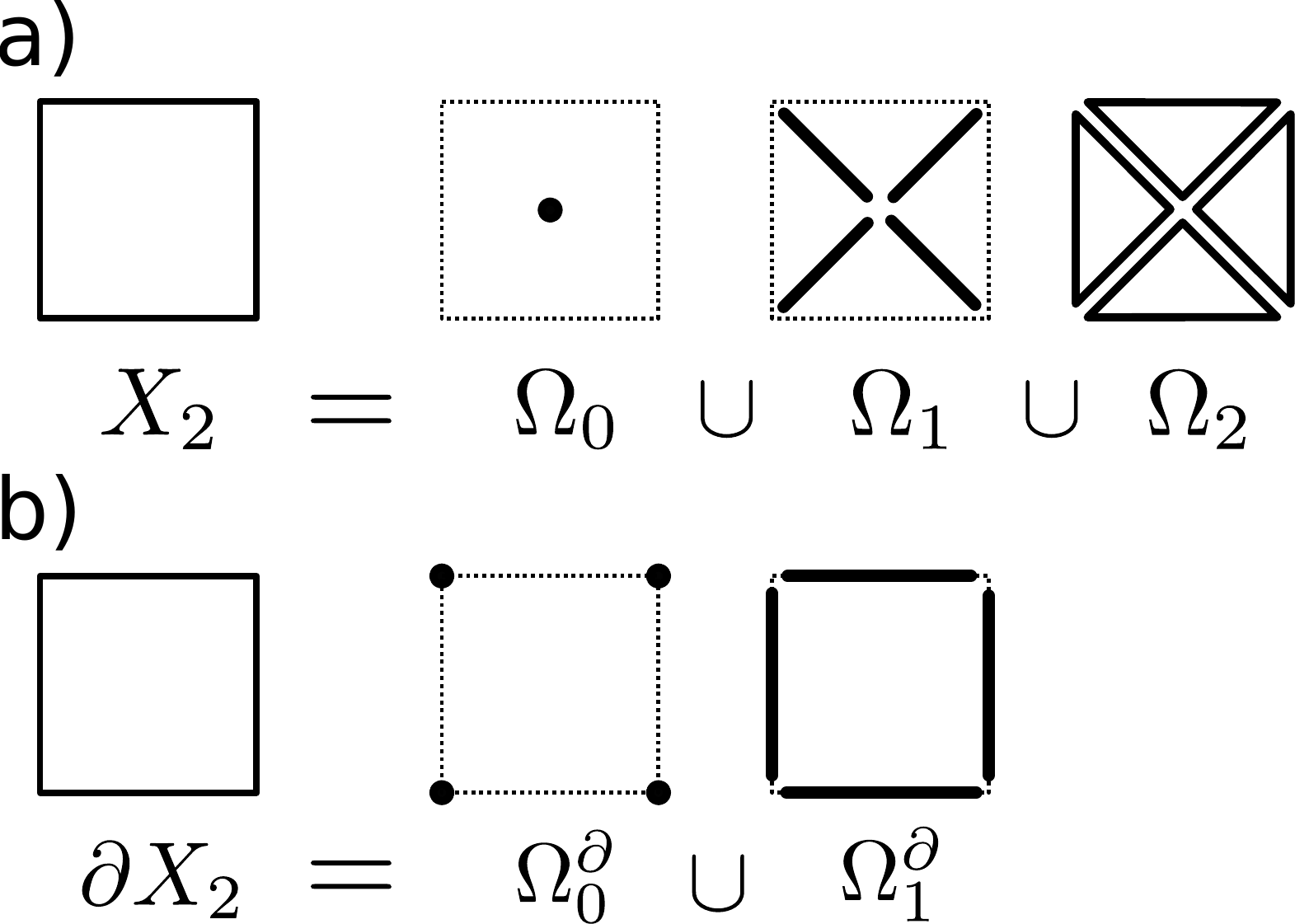}
		\caption{$G$-symmetric celular decomposition for $G=C_4$ (a). Induced boundary cellular decomposition (b).}
		\label{fig:bbh_decomp}
	\end{center}
\end{figure}

We first show that the boundary classification groups for this example are
\begin{equation}
  {\cal K}^{(1)}_{\rm a} = \ZZ,\ \
  {\cal K}^{(2)}_{\rm a} = \ZZ_2.\ \
  \label{eq:boundary_exampleapp}
\end{equation}
The ${\cal R}_4$ symmetric cellular decomposition is shown in
Fig.~\ref{fig:bbh_decomp}. First, we note that there can be no stable gapless points (Majorana fermions) at the $0$-cell $c_0\in\Omega_0$. To see this, note that since ${\cal R}_4$ 
acts onsite at $c_0$, any (Majorana) zero-energy bound states have well-defined angular momentum $j=1/2$, $3/2$, $5/2$, or $7/2$ (mod 4). Particle-hole conjugation pairwise connects these angular-momentum states, so that they can always gapped out. 

\begin{figure}[t]%
	\begin{center}
		\includegraphics*[width=0.9\columnwidth]{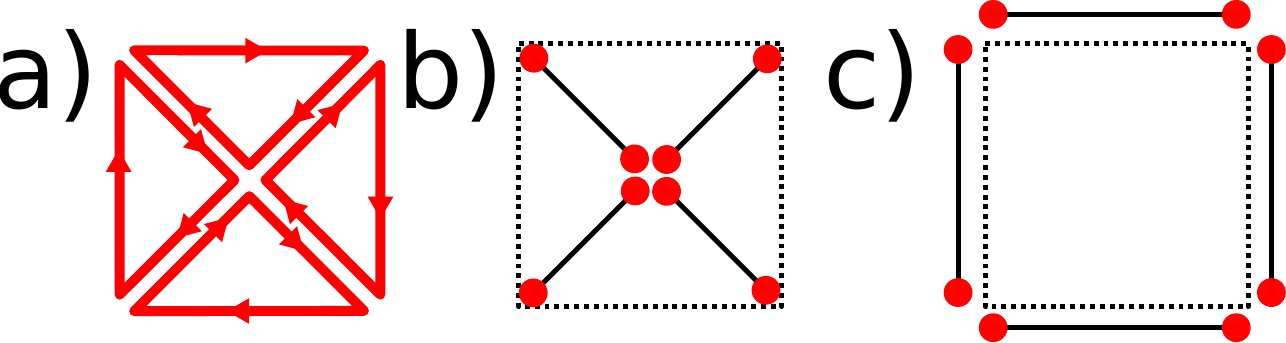}
		\caption{Classification of first-order phases by pasting two-dimensional Chern superconductors onto 2-cells from $\Omega_2$ (a). Second-order boundary states are classified by pasting one-dimensional superconductors onto 1-cells from $\Omega_1$ (b). Extrinsic second-order states are classified by pasting one-dimensional superconductors onto boundary 1-cells from $\Omega_1^\partial$.}
		\label{fig:bbh_app}
	\end{center}
\end{figure}

To classify first-order phases, we place two-dimensional Chern superconductors
onto the four 2-cells from $\Omega_2$, see Fig.~\ref{fig:bbh_app}a. The
counter-propagating chiral Majorana modes can be gapped out on 1-cells from
$\Omega_1$, whereas, as shown above, the 0-cell from $\Omega_0$ does not support stable zero-energy states. Since Chern superconductors have a $\ZZ$ classification, we obtain ${\cal
K}^{(1)}_{\rm a} = {\cal K}^{(1)} = \ZZ$.

Anomalous second-order phases are classified by placing one-dimensional Kitaev
chains onto 1-cells from $\Omega_1$ and using the fact that $\Omega_0$ does
not support topologically protected Majorana zero-states, see
Fig.~\ref{fig:bbh_app}b. This gives us ${\cal K}^{(2)} = \ZZ_2$. Extrinic
states are obtained by placing Kitaev chains onto bounday 1-cells from
$\Omega_1^\partial$, see Fig.~\ref{fig:bbh_app}c; Since all boundary states
obtain this way can be gapped out, it follows ${\cal D}^{(2)}=0$ and, hence, ${\cal
K}^{(2)}_{\rm a} = \ZZ_2$.

\subsection{Bulk classification}
The general method outlined in Sec.~\ref{sec:4.2} gives the bulk subgroup
sequence for two-dimensional odd-parity superconductor with point group $G = C_{4}$ 
\begin{equation}
  2\ZZ\times\ZZ \subset \ZZ^2 \subset \ZZ^3,
\end{equation}
which is consistent with the boundary
classification~(\ref{eq:boundary_exampleapp}). To obtain the bulk
classification group $K$, we first classify Dirac Hamiltonians
\begin{align}
	H&=m\Gamma_0+k_1\Gamma_1+k_2\Gamma_2,
	\label{eq:2dirac}
\end{align}
subject to the antiunitary antisymmetry ${\cal P}$ with ${\cal P}^2=1$ and to
the fourfold rotation symmetry ${\cal R}_4$, for which with the representation
$U_{\cal R}$ satisfies $U_{\cal R}^4=-1$ and commutes with ${\cal P}$. To apply
the Cornfeld-Chapman isomorphsm, we construct an additional fourfold symmetry
representation $U_{\cal R}^\Gamma=e^{\Gamma_1\Gamma_2\pi/4}$ from the Dirac
gamma matrices. The Cornfeld-Chapman isomorphism then maps the fourfold
rotation symmetry ${\cal R}_4$ to the fourfold local symmetry
\begin{align}
	U_{\cal O}&=U_{\cal R}^{\Gamma}U_{\cal R}.
	\label{eq:U4local}
\end{align}
The local symmetry $U_{\cal O}$ commutes with ${\cal P}$ and satisfies $U_{\cal
O}^4=1$. Considering the basis with well defined angular momentum $j=0$, $1$,
$2$, $3$ (mod 4), the Hamiltonian~(\ref{eq:2dirac}) takes block-diagonal form
$H=\mathrm{diag}(h_0,h_1,h_2,h_3)$. The blocks $h_j$ with even angular momentum
are subjected to particle-hole constraint ${\cal P}$ and belong to tenfold-way
class D, so that they have a $\ZZ$ classification. On the other hand, the
blocks $h_1$ and $h_3$ are related by particle-hole conjugation, hence for
their classification we need to consider the block $h_1$ only. Since this block
does not have any constraints, it belongs to class A and has a $\ZZ$
classification in two spatial dimensions. We thus conclude that $K=\ZZ^3$. The
topological invariants are $(\mbox{Ch}_0,\mbox{Ch}_1,\mbox{Ch}_{2})$ where
$\mbox{Ch}_j$ is the Chern number corresponding to the block with angular
momentum $j$,
\begin{align}
	\mbox{Ch}_j&=\frac{1}{2}\mathrm{tr}[\gamma_0^{(j)}\gamma_1^{(j)}\gamma_2^{(j)}],\ \ j=1,2,3,
	\label{eq:chern}
\end{align}
where the Hamiltonian $h_j$ is written in the form~(\ref{eq:2dirac}) with
$\Gamma_i$ matrices $\gamma_i^{(j)}$.

To find the group $K^{(2)}$, we calculate the classifying group $K_{O_2}$ of
Dirac Hamiltonians with two defect cordinates $x$ and $y$,
\begin{align}
	H_{O_2}&=H+x M_1+y M_2.
	\label{eq:h02}
\end{align}
Similarly as before, we use the Gamma matrices to construct the representation of the fourfold rotation symmetry
$\tilde U^\Gamma_{\cal R}=e^{\Gamma_1\Gamma_2\pi/4}e^{M_1M_2\pi/4}$. The
corresponding local symmetry reads
\begin{align}
	\tilde U_{\cal O}&=\tilde U_{\cal R}^\Gamma U_{\cal R},
	\label{eq:U4Rlocal}
\end{align}
which commutes with ${\cal P}$ and satisfies $\tilde U_{\cal O}^4=-1$. In the
basis with well defined angular momentum $j=1/2$, $3/$, $5/2$ ,$7/2$ (mod 4), the
Hamiltonian~(\ref{eq:h02}) takes block-diagonal form
$H_{O_2}=\mathrm{diag}(h_{1/2},h_{3/2},h_{5/2},h_{7/2})$. Particle-hole
symmetry relates the blocks $h_j$ and $h_{-j}$ to each other, thus we need to classify the blocks
$h_{1/2}$ and $h_{3/2}$ only. These two blocks have no symmetry constraints and
belong to class A. The classification of two-dimensional Hamiltonians with two
defect dimensions is isomorphic to the classification of zero-dimensional
Hamiltonians~\cite{teo2010}, i.e., $K_{O_2}=\ZZ^2$.

It remains to find the image of the inclusion $K^{(2)} \hookrightarrow K$.
Hereto, we calculate the topological invariants $\mbox{Ch}_0$, $\mbox{Ch}_1$
and $\mbox{Ch}_2$ for the two generators of $K_{O_2}$. We represesnt these
generator by Hamiltonians of the form (\ref{eq:h02}), with Gamma matrices and
mass terms given by
\begin{align}
	\tilde\gamma_0&=-\tau_3\mu_3, \nonumber \\
	(\tilde\gamma_1,\tilde\gamma_2)&=(\tau_3\mu_3\sigma_2,\tau_3\mu_2), \nonumber \\
	(\tilde m_1,\tilde m_2)&=(\tau_3\mu_3\sigma_1,\tau_3\mu_1),
	\label{eq:C4gen}
\end{align}
and with the representations $U_{\cal P}=\tau_1$ and $\tilde u_{\cal O}=e^{i\tau_3\pi/4}$, $e^{3 i \tau_3 \pi/4}$ for 
particle-hole conjugation and for the on-site fourfold 
symmetry for the two generators, respectively. (The only
difference between the two generators is the representation of $\tilde u_{\cal
O}$.) Using Eq.~(\ref{eq:U4Rlocal}), we find that the original fourfold
rotation symmetry has the representation $U_{\cal
R}=e^{i\mu_2\sigma_1\pi/4}e^{i\mu_2\sigma_2\pi/4}\tilde u_{\cal O}$. The
corresponding local
symmetry $U_{\cal O}$ then reads
\begin{align}
	U_{\cal O}&=e^{i\mu_2\sigma_1\pi/4} \tilde u_{\cal O}.
	\label{eq:U4}
\end{align}
For $\tilde u_{\cal O}=e^{i\tau_3\pi/4}$ there are four states with angular momentum
$j=0$ and four states with odd angular momentum. Hence, $\mbox{Ch}_{2}=0$, and
explicit calculation gives $\mbox{Ch}_0=-2 \mbox{Ch}_1$ and $\mbox{Ch}_1=1$.
Similarly, for the generator with $\tilde u_{\cal O}=e^{i\tau_3 3\pi/4}$ we
find that $\mbox{Ch}_1=-2 \mbox{Ch}_{2}$ and $\mbox{Ch}_{2}=1$. Hence, the
image of $K_{O_2}\hookrightarrow K$ is generated by the elements $(2,-1,0)$ and
$(0,-1,2)$. As a subgroup of $K$, the image of $K_{O_2}\hookrightarrow K$ is
isomorphic to $\ZZ \times 2 \ZZ$.

In order to find the subgroup $K^{(1)}$ one can proceed in two ways: either one
classifies Dirac Hamiltonians with a single mass term transforming under the one-dimensional
representation $O_1({\cal R}_4)=-1$ or one finds the kernel of the
homomorphism $K\hookrightarrow K_\text{TF}$ where the group $K_\text{TF}=\ZZ$ classifies two-dimensional phases in class D without additional constraints. It
is easy to see that this homomorphism acts as
\begin{align}
	(\mbox{Ch}_0,\mbox{Ch}_1,\mbox{Ch}_{2})\rightarrow \mbox{Ch}_0+2 \mbox{Ch}_1+\mbox{Ch}_{2},
	\label{eq:kernel}
\end{align}
hence $K^{(1)}=\ZZ^2$.

For illustration purposes, we also calculate $K^{(1)}$ using the alternative
approach, i.e., by considering the one-dimensional representation $O_1({\cal
R}_4)=-1$. The group $K_{O_1}$ classifies the Hamiltonians 
\begin{align}
	H_{O_1}&=H+(x^2-y^2)M.
	\label{eq:h01}
\end{align}
Altough the above Hamiltonian is not in Dirac form, we find the representation
of the fourfold rotation antisymmetry $\bar U_{\cal
CR}^\Gamma=e^{\Gamma_1\Gamma_2\pi/4}M$. Using the Cornfeld-Chapman isomorphism we
classify Hamiltonians of the form of Eq.~(\ref{eq:h01}) with onsite antisymmetry
\begin{align}
	\bar U_{\cal R}^{\cal C}&=\bar U^\Gamma_{\cal CR} U_{\cal R},
\end{align}
where $(\bar U_{\cal R}^{\cal C})^4=1$. The twofold rotation symmetry $U_{{\cal
R}_2}^{\cal O}=(U_{\cal R}^{\cal C})^2$ commutes with ${\cal P}$, and states
with $U_{{\cal R}_2}^{\cal O}=\pm1$ define two blocks
$H_{O_1}=\mathrm{diag}(h_+,h_{-})$. The block $h_+$ has chiral constraint $\bar
U_{\cal R}^{\cal C}$ that squares to one, thus it belongs to class DIII.
Similarly, we find that the block $h_-$ belongs to class BDI. Therefore,
$K_{O_1}=\ZZ\times\ZZ_2$. The $\ZZ_2$ part of $K_{O_1}$ has to be in the kernel
of the inclusion $K_{O_1}\hookrightarrow K=\ZZ^3$. We only need to consider the
generator of the free part of the group $K_{O_1}$. The generator Hamiltonian
reads

\begin{align}
	\bar\gamma_0&=\tau_3\sigma_2, \nonumber \\
	(\bar\gamma_1,\bar\gamma_2)&=(\tau_3\sigma_1,\tau_3\sigma_3), \nonumber \\
	\bar m&=\tau_2,
\end{align}
with symmetries $\bar u_{\cal R}^{\cal C}=i\tau_1$, $U_{\cal P}=1$. The
representation of the crystalline symmetry is $U_{\cal
R}=e^{i\sigma_2\pi/4}\tau_3$. The local symmetry representation is found to be
$U_{\cal O}=\tau_3$. Thus the generator has topological invariants
$\mbox{Ch}_0=\mbox{Ch}_{2}=1$ and $\mbox{Ch}_1=0$. The product of the subgroup $K^{(2)}\subseteq K$
and the subgroup of the elements $\mbox{Ch}_0=-\mbox{Ch}_1$, $\mbox{Ch}_1=0$ gives the
subgroup $K^{(1)}$ previously found.

\bibliography{refs}
\end{document}